\documentclass[preprint,prb]{revtex4}
\usepackage[T1]{fontenc}
\usepackage{graphicx}
\usepackage{rotating}
\usepackage{bm}        
\usepackage{amsmath}
\usepackage{amssymb}   
\usepackage{bm}
\def\jcp#1#2#3{J.~Chem.~Phys.~{\bf #1},\ #2\ (#3)}
\def\cpl#1#2#3{Chem.~Phys.~Lett.~{\bf #1},\ #2\ (#3)}

\def\pra#1#2#3{Phys.~Rev.~A~{\bf #1},\ #2\ (#3)}
\def\prl#1#2#3{Phys.~Rev.~Lett.~{\bf #1},\ #2\ (#3)}

\def\jpb#1#2#3{J. Phys. B: At. Mol. Opt. Phys. {\bf #1},\ #2\ (#3)}

\def\hr{\hat{r}_\alpha}
\def\hR{\hat{R}_\alpha}

\def\sr{\hat{r}}
\def\sR{\hat{R}}

\def\br{{\bf r}_\alpha}
\def\bR{{\bf R}_\alpha}

\def\k1{k_1}
\def\k2{k_2}
\def\q1{q_1}
\def\q2{q_2}

\def\({\left (}
\def\){\right )}
\def\[{\left [}
\def\]{\right ]}

\newcommand{\beq}{\begin{equation}}
\newcommand{\eeq}{\end{equation}}

\newcommand{\threejm}[6]{ \left(\begin{array}{ccc} #1 & #3 & #5\\
                                              #2 & #4 & #6
                                \end{array}
                          \right)}

\renewcommand{\l}{\ell}

\begin{document}
\date{\today}
\title{
Quantum theory of chemical reactions in the presence of electromagnetic fields}
\author{T. V. Tscherbul}
\email{timur@chem.ubc.ca}
\author{R. V. Krems}
\affiliation{Department of Chemistry, University of British Columbia, Vancouver, B.C. V6T 1Z1, Canada}
\begin{abstract}
We present a theory for rigorous quantum scattering calculations of probabilities for chemical reactions of atoms with diatomic molecules in the presence of an external electric field. The approach is based on the fully uncoupled basis set representation of the total wave function in the space-fixed
coordinate frame, the Fock-Delves hyperspherical coordinates and adiabatic partitioning of the total Hamiltonian of the reactive system. The adiabatic channel wave functions are expanded in basis sets of hyperangular functions corresponding to different reaction arrangements and the interactions with external fields are included in each chemical arrangement separately. We apply the theory to examine the effects of electric fields
on the chemical reactions of LiF molecules with H atoms and HF molecules with Li atoms at low temperatures
and show that electric fields may enhance the probability of chemical reactions and modify reactive scattering resonances by coupling the
rotational states of the reactants.
Our preliminary results suggest that chemical reactions of polar molecules at temperatures below 1 K can be
selectively manipulated  with dc electric fields and microwave laser radiation.

\end{abstract}

\maketitle

\clearpage
\newpage

\section{Introduction}

An important goal of modern chemical physics is to achieve external control over dynamics of elementary chemical processes
\cite{Zare,Roman,LoeschReview,Stereodynamics,EH,Moshe,Bretislav}. Manipulating chemical transformations by external dc fields
or laser radiation is at the heart of mode-selective chemistry \cite{Zare},
chemical stereodynamics \cite{LoeschReview,Stereodynamics} and quantum coherent control of molecular dynamics \cite{Moshe}.
External electromagnetic fields can be used to orient and align molecules, which restricts
the symmetry of the electronic interaction between the reactants in the entrance reaction channel and may result in suppression or
enhancement of reaction rates, the phenomenon known as the ``steric effect'' \cite{Steric1,Brooks,Stolte}. 
Loesch and co-workers \cite{LoeschReview,Loesch2,LoeschMoller}
and Friedrich and Herschbach \cite{Bretislav} demonstrated that rotationally cold polar molecules in the $\Sigma$ electronic state
can be effectively oriented by dc electric fields which was used to study steric effects in molecular spectroscopy \cite{Bretislav1},
inelastic scattering \cite{Bretislav}, and chemical reaction dynamics \cite{LoeschReview,LoeschMoller}.
Loesch and Stienkemeier used a combination of dc electric fields and infrared radiation pumping to explore the
effects of molecular alignment in the Li~+~HF$(v=1,j=1)$ chemical reaction. Their results indicated that side-on collisions between HF molecules and Li atoms
are more likely to result in the reaction than end-on collisions \cite{LoeschMoller}.
The steric effects observed in experiments with thermal molecular beams are, however, usually weak \cite{LoeschReview}
because the kinetic energy of the reactants greatly exceeds the perturbations induced by dc electric fields, even for very
polar and heavy molecules like ICl \cite{LoeschMoller}.

Friedrich and Herschbach have shown that molecules can also be aligned by laser radiation \cite{FH1995}.
The laser alignment method can be applied to both polar and non-polar molecules \cite{Seideman}. Stapelfeldt
and co-workers \cite{Stapelfeldt} demonstrated that significant alignment can be achieved with laser fields of
$10^{12}$ W/cm$^2$. The degree of alignment can be quantified by photoionizing the aligned
molecules and examining the angular distributions of the photofragments \cite{Seideman,Stapelfeldt}. Laser-field alignment
has been used to produce high-order harmonics with specific polarization emitted
by N$_2$, O$_2$, and CO$_2$ molecules \cite{Corkum}.
Laser-field alignment can also be used to manipulate the rotational motion of molecules \cite{Stapelfeldt2} or control
the branching ratios of the photodissociation products \cite{Stapelfeldt3}. 
The interaction of molecules with an off-resonant laser light is proportional to the square of the electric field strength
and substantial alignment can be achieved only with very powerful lasers.
Because most lasers have short duty cycles, laser-aligned molecules are normally produced with low densities insufficient for
scattering experiments   \cite{Stereodynamics,EH}.
Other methods, such as collisional alignment in supersonic expansions \cite{Stereodynamics}, produce large quantities of aligned
molecules, but the degree of alignment in these experiments \cite{CollisionalA} is often insignificant and difficult to quantify.

The effects of external fields on molecular collisions are significantly more pronounced at low temperatures. The development of experimental
techniques for cooling molecules to temperatures near or below 1 K has opened up new possibilities to study controlled chemical reactions \cite{RomanPCCP}.
Chemical reactions of molecules at cold and ultracold temperatures are accelerated by resonances \cite{BalaReview,BalaLiF},
tunneling \cite{Roman,BalaReview,BalaLiF}, threshold phenomena \cite{UltracoldChemistry}, quantum interference \cite{SuperChemistry} and many-body
dynamics \cite{SuperChemistry,NatureInsight}.
With the development of novel experimental methods for manipulating molecules with electromagnetic fields
such as Stark deceleration \cite{StarkDeceleration,MeijerXeOH}, magnetic or electrostatic guiding \cite{Rempe} and the design of
a molecular synchrotron \cite{MeijerMS}, it has become possible to study cold
chemical reactions in the presence of external fields experimentally \cite{Roman,NNH}.
Staanum {\it et al.} \cite{Alkali1} and Zahzam {\it et al.} \cite{Alkali2} have recently reported measurements of
inelastic collisions and chemical reactions in an optically
trapped mixture of Cs$_2$ molecules and Cs atoms. Several research groups are currently developing experiments
to study chemical reactions of formaldehyde with OH radicals \cite{HudsonFormaldehyde}, and Li atoms with HF molecules
\cite{HeraeusSeminar} in slow molecular beams.
In order to interpret the experimental data and identify new directions for research with cold molecules, it is necessary to develop rigorous
scattering theory of low-temperature chemical reactions in the presence of external fields.

The first theoretical studies of chemical reactions in the presence of external dc fields date back to the work of
Karplus and Goldfrey \cite{Steric2}, who used quasiclassical trajectory (QCT) calculations based on two different potential
energy surfaces to explain the experimental observations for the Rb + CH$_3$I reaction \cite{Steric1}.
More recently,  Aoiz and co-workers \cite{BretislavQCT} used QCT 
simulations to explore the effects of electric fields on the reaction DCl~+~H $\to$ HCl~+~D at thermal energies.
Aldegunde {\it et al.}  \cite{Aldegunde1,Aldegunde2} proposed to describe alignment of the reactants in terms of polarization moments
to examine steric effects in the chemical reactions of H with D$_2$ and F with H$_2$.
Using time-independent reactive scattering calculations,
they demonstrated that the differential scattering cross sections for the F + H$_2$ chemical reaction at ultralow temperatures
can be controlled by changing the polarization of the H$_2$ molecules \cite{Aldegunde2}. 
Aquilanti and co-workers  \cite{Enzo} developed a stereodirected representation for the scattering $S$-matrix 
to analyze steric effects in the Li + HF reaction in the absence of external fields.

Model theoretical studies of chemical reactions in laser fields  
have been reported by many groups \cite{OM,Ivanov,Light,Moshe1,Moshe2,Moshe3,TannorRice}. Orel and Miller
found that the collinear chemical reactions of H, F, and Cl atoms with H$_2$ molecules can be enhanced by intense off-resonant laser fields.
Altenberger-Siczek and Light \cite{Light} used the Floquet formalism to study the collinear reaction LiF~+~H $\to$ LiF~+~H  in an optical field. 
Orel and Miller \cite{OM} reported QCT calculations for the same system. Both studies indicated that the reaction probability may be enhanced
in the presence of laser fields. The enhancement is a result of a laser-induced avoided crossing between the ground and the
first excited electronic states, reducing the reaction barrier.
Seideman and Shapiro \cite{Moshe1} developed an approximate theory of laser catalysis for chemical reactions in 3D. They demonstrated that the
chemical reactions H~+~H$_2$ \cite{Moshe1} and D + H$_2$ \cite{Moshe2} can be controlled by coupling two
electronic states with a picosecond laser pulse. Li {\it et al.} \cite{Moshe3} have recently applied a time-dependent wave packet method
to study laser catalysis of the Li$_2$~+~Li exchange reaction at low temperatures.
Tannor and Rice \cite{TannorRice} proposed a mechanism to selectively control chemical reactions with an optimized sequence
of two femtosecond laser pulses. All of the above studies,
however, relied on significant approximations to simplify the reactive scattering problem in the presence of external fields. For example, the effects
of symmetry breaking in strong laser fields were not properly considered.
These approximations may not be adequate for dynamics of chemical reactions at low temperatures.
As the collision energy of molecules in the subKelvin temperature regime is usually smaller than the perturbations due to external fields,
the scattering theory of cold chemical reactions must explicitly include the interactions with external fields \cite{Roman}.

 The purpose of the present paper is to develop a rigorous quantum theory of abstraction chemical reactions in the presence of an electric field.
 Our formalism is based on the fully uncoupled space-fixed (SF) basis representation of the total wave function expressed in the Fock-Delves
 hyperspherical coordinates to describe the reactive scattering problem.
Rigorous quantum theory of non-reactive molecular collisions in external electric and magnetic fields was
initially developed by Volpi and Bohn \cite{Volpi} and Krems and Dalgarno \cite{Roman2004,OurWork,DCS}.
These authors demonstrated that inelastic collisions of molecules at low energies can be manipulated
with electromagnetic fields.  Here, we generalize the work of Krems and Dalgarno to describe
chemical reactions in external fields.  Our calculations indicate that
the probability of the LiF~+~H $\to$ HF + Li chemical reaction is sensitive to electric fields of less than 200~kV/cm.
The remainder of this paper is organized as follows. Section II presents the theory. In Sec.~III, we  examine the effects of electric
fields on the cross sections and rate constants for the Li + HF($v=0,j=0$) $\to$ LiF + H and LiF($v=1,j=0$)~+~H $\to$ HF~+~Li chemical reactions
at low temperatures. Sec. IV summarizes possible applications of our theory and outlines future prospects.

\section{Theory}

\subsection{Hamiltonian and coordinates}\label{SIIA}

The Hamiltonian for an atom-diatomic molecule collision system \cite{ParkerPack,Review} in the presence of
a homogeneous dc electric field can be written in atomic units as 
\begin{align}\label{H}\notag
H=-\frac{1}{2\mu R_\alpha} \frac{\partial^2}{\partial R_\alpha^2}R_\alpha
+\frac{\bm{\ell}^2_\alpha}{2\mu R_\alpha^2} &+ V(R_\alpha,\,r_\alpha,\,\gamma_\alpha) -V_\alpha(r_\alpha) 
\\ &- [{\bf d}(R_\alpha,\,r_\alpha,\,\gamma_\alpha) - {\bf d}_\alpha(r_\alpha)]\cdot {\bf E} + H_\text{as},
\end{align}
where $\bR$ and $\br$ are the mass-scaled Jacobi coordinates \cite{Review}, $\mu$ is the
three-body reduced mass, $\gamma_\alpha$ is the angle between the vectors $\bR$ and $\br$,
$V(R_\alpha,\,r_\alpha,\,\gamma_\alpha)$ is the interaction potential, and
$\bm{\ell}_\alpha$ is the orbital angular momentum describing the mechanical rotation of the
reactive complex. In Eq. (\ref{H}),
${\bf d}(R_\alpha,\,r_\alpha,\,\gamma_\alpha)$ is the dipole moment of the triatomic system, and ${\bf E}$ is the electric field vector 
which defines the space-fixed quantization axis $z$. We note that
\begin{equation}\label{Dlimit}
\lim_{R_\alpha \to \infty} {\bf d}(R_\alpha,\,r_\alpha,\,\gamma_\alpha) = {\bf d}_\alpha(r_\alpha),
\end{equation}
where ${\bf d}_\alpha(r_\alpha)$ is the permanent dipole moment of the diatomic molecule.
The subscript $\alpha$ in Eqs. (\ref{H}) and (\ref{Dlimit}) refers to different chemical arrangements \cite{ParkerPack,Review}.
The asymptotic Hamiltonian describes an isolated $^1\Sigma$ diatomic molecule in the presence
of an electric field \cite{LandauLifshitz,OurWork}
\begin{equation}\label{HmolJacobi}
H_\text{as} = H_\text{mol} + H_\text{ef}
\end{equation}
where
\begin{equation}\label{HmolJacobi1}
H_\text{mol} = -\frac{1}{2\mu r_\alpha}\frac{\partial^2}{\partial r_\alpha^2}r_\alpha
+ \frac{{\bf j}^2_\alpha}{2\mu r_\alpha^2}+V(r_\alpha),
\end{equation}
and
\begin{equation}\label{HmolJacobi2}
H_\text{ef} = - {\bf d}_\alpha(r_\alpha)\cdot {\bf E}.
\end{equation}
Here, $V(r_\alpha)$ is the potential energy function, ${\bf j}_\alpha$ is the rotational angular momentum of the diatomic molecule,
 and $H_\text{ef}$ describes the interaction of the molecule with the electric field.

The reactive scattering problem is most conveniently formulated in the Fock-Delves (FD) hyperspherical
coordinates - the hyperradius $\rho$, the hyperangle $\theta_\alpha$,
and the Jacobi angle $\gamma_\alpha$ - related to the mass-scaled Jacobi coordinates as follows
\begin{align}\label{FD}\notag
r_\alpha &= \rho\sin\theta_\alpha, \\
R_\alpha &= \rho\cos\theta_\alpha.
\end{align}
The Hamiltonian (\ref{H}) can be expressed in the FD coordinates \cite{ParkerPack,Review}
\begin{align}\label{HFD}\notag
\tilde{H}=-\frac{1}{2\mu\rho^5}\frac{\partial}{\partial \rho}\rho^5\frac{\partial}{\partial \rho}
+ \frac{\bm{\ell}^2_\alpha}{2\mu\rho^2\cos^2\theta_\alpha}
&+V(\rho,\,\theta_\alpha,\,\gamma_\alpha) - V_{\alpha}(r_{\alpha})
\\ &- [{\bf d}(\rho,\,\theta_\alpha,\,\gamma_\alpha) - {\bf d}_\alpha(\rho,\theta_\alpha)]\cdot {\bf E} + \tilde{H}_\text{as}.
\end{align}
The FD asymptotic Hamiltonian  $\tilde{H}_\text{as}$ can be represented as
\begin{equation}\label{Hmol}
\tilde{H}_\text{as}= \tilde{H}_\text{mol} + \tilde{H}_\text{ef},
\end{equation}
where the operators
\begin{equation}\label{Hmol1}
\tilde{H}_\text{mol} = 
\frac{1}{2\mu\rho^2}\left[
-\frac{1}{\sin^22\theta_\alpha}\frac{\partial}{\partial \theta_\alpha}\sin^22\theta_\alpha
\frac{\partial}{\partial \theta_\alpha}+ \frac{{\bf j}^2_\alpha}{\sin^2\theta_\alpha}\right]
+V_\alpha(r_\alpha)
\end{equation}
and
\begin{equation}\label{Hef}
 \tilde{H}_\text{ef} =- {\bf d}_\alpha(\rho,\theta_\alpha)\cdot {\bf E}
\end{equation}
are obtained by the coordinate transformation from Eqs. (\ref{HmolJacobi1}) and (\ref{HmolJacobi2}).
We use the tilde to denote the operators and functions expressed in the FD coordinates. 

The total wave function of the reactive complex can be expanded as \cite{ParkerPack,Review,Tolstikhin}
\begin{equation}\label{TotalWF}
\Psi = \rho^{-5/2}\sum_i F_i(\rho)\Phi_i({\omega;\rho}),
\end{equation}
where $\omega$ denotes collectively the hyperangles and $\rho$ is the hyperradius defined by Eq. (\ref{FD}). 
We emphasize that Eq. (\ref{TotalWF}) is written for a particular component of the wave function
$\Psi_{i_0}$, where $i_0$ labels the molecular states before the collision. It is therefore important to
remember that the expansion coefficients in Eq. (\ref{TotalWF}) depend on two indices $i$ and $i_0$, which
is often written as $F_{ii_0}$ or $F^{i_0}_i$. Each column of the matrix of the coefficients $F$ is a linearly independent solution of the
Schr{\"o}dinger equation corresponding to the initial state $i_0$ \cite{ParkerPack,Review}
\begin{equation}\label{SE}
\tilde{H}\Psi_{i_0} = E_\text{t}\Psi_{i_0},
\end{equation}
where $E_\text{t}$ is the total energy.
We will use the index $i_0$ only when necessary.

The basis functions $\Phi_i({\omega;\rho})$ can be chosen as the eigenfunctions of the adiabatic Hamiltonian \cite{Review,Tolstikhin}
\begin{equation}\label{AEP}
\tilde{H}_\text{ad}(\rho)\Phi_i({\omega;\rho}) = \epsilon_i(\rho)\Phi_i({\omega;\rho}),
\end{equation}
where $\epsilon_i(\rho)$ are the adiabatic eigenvalues, which depend parametrically on $\rho$.
In this work, we define the adiabatic Hamiltonian as the total Hamiltonian (\ref{HFD}) without the hyperradial kinetic energy
\begin{equation}\label{Had}
\tilde{H}_\text{ad}(\rho) = \tilde{H}_\text{fi}(\rho) + \tilde{H}_\text{ef},
\end{equation}
where $ \tilde{H}_\text{ef}$ is the Hamiltonian for the molecule-field interaction given by Eq. (\ref{Hef}) and the
field-independent term can be written in the form
\begin{equation}\label{Hfi}
 \tilde{H}_\text{fi}(\rho) =  \frac{\bm{\ell}^2_\alpha}{2\mu\rho^2\cos^2\theta_\alpha}
+V(\rho,\,\theta_\alpha,\,\gamma_\alpha) - V_{\alpha}(r_{\alpha}) +\tilde{H}_\text{mol}.
\end{equation}
In deriving Eqs. (\ref{Had}) and (\ref{Hfi}), we have neglected the interaction of the three-body component of the dipole moment with electric fields
\begin{equation}\label{H3b}
\tilde{H}_\text{ef,3B} = - [{\bf d}(\rho,\,\theta_\alpha,\,\gamma_\alpha) - {\bf d}_\alpha(\rho,\theta_\alpha)]\cdot {\bf E}.
\end{equation}
Ab initio calculations show that the expression in the square brackets decreases rapidly with increasing $\rho$ \cite{Truhlar}.
At small $\rho$, the interaction of a transient reaction complex with electric fields is negligible in comparison with
the electronic interaction potential of the complex \cite{Vigue}. 
Since the interaction due to the three-body term in Eq. (\ref{H3b}) couples the same states as the interaction potential,
it can be safely neglected  \cite{Vigue}. It was demonstrated in our previous work \cite{Roman2006}
that including the interaction with the transient dipole moment in scattering calculations does not modify
the collision dynamics except near certain scattering resonances at extremely low temperatures. 

Substituting the expansion (\ref{TotalWF}) into the Schr{\"o}dinger equation with the Hamiltonian (\ref{HFD}) and
using Eq. (\ref{AEP}), we obtain a system of close-coupled (CC) equations \cite{ParkerPack,Review,David,HuSchatz,NymanYu}
\begin{equation}\label{CC}
\biggl{[} \frac{d^2}{d\rho^2} - \frac{15}{8\mu \rho^2} + 2\mu[E-\epsilon_i(\rho)]\biggr{]}F_i(\rho) = 0.
\end{equation}
Eqs. (\ref{TotalWF}), (\ref{AEP}) and (\ref{CC}) can be solved on a grid of radial sectors $\rho_k$ extending from small values
of $\rho$ to $\rho = \infty$ to determine the total wave function of the reactive complex subject to the scattering boundary conditions.
The reactive scattering problem can thus
be separated into three steps: (i) the diagonalization of the adiabatic Schr\"odinger equation (\ref{AEP});
(ii) the integration of the CC equations (\ref{CC}), and (iii) matching the results of the numerical integration to
the asymptotic form of the wave function as determined by reactive scattering boundary conditions.

\subsection{The adiabatic eigenvalue problem}

The adiabatic eigenvalue problem (\ref{AEP}) can be solved using an expansion
\begin{equation}\label{Expan}
\Phi_i({\omega;\rho}) = \sum_{n}T_{ni}|\zeta_n\rangle,
\end{equation}
where $|\zeta_n\rangle$ are some orthonormal basis functions.
Substituting this expansion into Eq. (\ref{AEP}) leads to a matrix eigenvalue problem
\begin{equation}\label{EVP}
\sum_{n'}\bigl{[}
\langle \zeta_n | H_\text{ad}(\rho) | \zeta_{n'}\rangle - \epsilon_i(\rho)\delta_{nn'}\bigr{]}T_{n'i} = 0.
\end{equation}
The eigenvalues $\epsilon_i(\rho)$ and eigenvectors $T_{ni}$ can be found once the adiabatic Hamiltonian
matrix is specified in the basis $|\zeta_i\rangle$. In order to construct this matrix, we introduce a
basis set that simultaneously includes primitive functions defined in all chemical arrangements \cite{Miller,AMW}
\begin{equation}\label{primitive}
 \chi_{\alpha vj}(\theta_\alpha;\rho) |jM_j\rangle |\ell M_\ell\rangle = \chi_{\alpha vj}(\theta_\alpha;\rho)Y_{jM_j}(\hr) Y_{\ell M_\ell}(\hR)
\end{equation}
where the uncoupled space-fixed (SF) basis functions are direct products of the spherical harmonics
$Y_{jM_j}(\hr)$ and $Y_{\ell M_\ell}(\hR)$, and the FD rovibrational eigenfunctions and eigenenergies are defined as
\begin{equation}\label{chis}
\tilde{H}_\text{mol}\chi_{\alpha vj}(\theta_\alpha;\rho) = \epsilon_{\alpha vj}(\rho)\chi_{\alpha vj}(\theta_\alpha;\rho).
\end{equation}
The projection of the total angular momentum onto the electric field axis $M=M_j+M_\ell$
is rigorously conserved for collisions in parallel fields \cite{Roman2004,Volpi}. Therefore, the basis functions
 (\ref{primitive}) corresponding to different values of $M$ are not coupled, and
the Schr{\"o}dinger equation (\ref{Had}) can be solved independently for each $M$.

We emphasize that the basis functions (\ref{primitive}) are not the eigenfunctions of the total
angular momentum operator and that the angular momentum projections $M_j$ and $M_\ell$ are defined
with respect to the SF quantization axis determined by the direction of the external field.
All previous methods for solving the reactive scattering problem 
used the total angular momentum representation of Arthurs and Dalgarno \cite{AD} in the body-fixed (BF) coordinate system.
The quantization axis in the BF approach is directed along the Jacobi
vector $\bR$. This choice simplifies the evaluation of the matrix elements of the interaction potential. In addition, chemical
reactions with near-collinear transition states are determined by a limited number of the BF projections of the total angular momentum
\cite{ABC}, which leads to a substantial reduction of the number of scattering channels.
External electric fields break the isotropy of space and couple states corresponding to different total angular momenta and inversion parities.
Therefore, the total angular momentum representation \cite{AD,Lester} offers no advantage. Krems and Dalgarno showed \cite{Roman2004}
that in the presence of an external field, it is more convenient to work directly in the SF frame as
this leads to simpler expressions for the matrix elements of the molecule-field interaction.
Although the matrix elements of the interaction potential in the SF basis are more complicated \cite{Roman2004},
they can be obtained from the BF matrix elements using a simple transformation as shown in Sec. II C below.

We note that the basis (\ref{primitive}) is not orthogonal because the functions
$\chi_{\alpha vj}(\theta_\alpha;\rho)$ of different $\alpha$ overlap at small $\rho$ \cite{Review}. 
An appropriate orthogonal basis set can be defined in terms of the functions (\ref{primitive}) following
the symmetric orthogonalization procedure \cite{ABC,Szabo,Lowdin} 
\begin{equation}\label{Zeta}
|\zeta_n\rangle = \frac{1}{\sqrt{\lambda_n}}\sum_{jM_j,\ell M_\l} {X}_{\alpha vjM_j\l M_\ell,n}
\chi_{\alpha vj}(\theta_\alpha;\rho) |jM_j\rangle |\ell M_\ell\rangle,
\end{equation}
where $\lambda_i$ and ${X}^{M}_{\alpha vjM_j \ell M_\ell,n}$ are, respectively, the real eigenvalues and eigenvectors
of the overlap matrix $\mathsf{O}$
\begin{equation}\label{Orthogonalization}
\mathsf{X}^\text{T}\mathsf{O}\mathsf{X} = \Lambda,
\end{equation}
where $\Lambda = \text{diag}(\lambda_1,\ldots,\lambda_n)$ and the symmetric and orthogonal
overlap matrix of the primitive basis functions (\ref{primitive}) is given by
\begin{equation}\label{metric}
\text{O}_{\alpha vjM_j\ell M_\ell,\alpha' v'j'M_j' \ell'M_\ell'} = \langle \chi_{\alpha vj}(\theta_\alpha;\rho)|\langle jM_j |\langle \ell M_\ell |
\chi_{\alpha'v'j'}(\theta_{\alpha'};\rho) \rangle |j' M_j' \rangle | \ell' M_\ell'\rangle.
\end{equation}

It is easy to verify that the functions defined by Eq. (\ref{Zeta}) are orthogonal,
and therefore can be used to expand the adiabatic surface functions in Eq. (\ref{AEP}).
The matrix elements of the adiabatic Hamiltonian (\ref{EVP}) can be obtained from Eq. (\ref{metric})
\begin{multline}\label{MatrixElementsHrho}
\langle \zeta_n | H_\text{ad}(\rho) | \zeta_{n'}\rangle = \frac{1}{\sqrt{\lambda_n\lambda_{n'}}}
\sum_{\alpha, v,j,M_j, \ell, M_\ell} \sum_{\alpha', v',j',M_j', \ell', M_\ell'}X_{\alpha vjM_j \ell M_\ell,n}
X_{\alpha' v'j'M_j'\ell' M_\ell',n'} \\ \times
\langle \chi_{\alpha vj}(\theta_\alpha;\rho)|\langle jM_j |\langle \ell M_\ell | H_\text{ad}(\rho)|
\chi_{\alpha'v'j'}(\theta_{\alpha'};\rho) \rangle |j' M_j' \rangle | \ell' M_\l'\rangle.
\end{multline}

Because the adiabatic functions calculated at different $\rho$ are not orthogonal, we
need an additional transformation between the adiabatic  functions of the adjacent propagation
sectors (say, $\rho_{k-1}$ and $\rho_k$). The transformation is derived in Sec. \ref{SIIE} and has the form
\begin{equation}\label{Sdefinition}
[\mathsf{S}(\rho_{k-1},\rho_{k})]_{ii'} = \langle \Phi_{i}(\omega;\rho_{k-1}) | \Phi_{i'}(\omega;\rho_k)\rangle,
\end{equation}
where the integration is carried out over all variables except $\rho$. Expanding the adiabatic surface functions
as in Eqs. (\ref{Expan}) and (\ref{Zeta}), we find
\begin{multline}\label{S}
[\mathsf{S}(\rho_{k-1},\rho_{k})]_{ii'} = \sum_{n,n'}T_{ni}T_{n'i'} 
\frac{1}{\sqrt{\lambda_n\lambda_{n'}}}\sum_{\alpha, v, j, M_j, \ell, M_\ell}\sum_{\alpha'v'j'M_j'\ell'M_\ell'}
X_{\alpha vjM_j\ell M_\ell,n}(\rho_{k-1}) X_{\alpha' v'j'M_j'\ell' M_\ell',n'}(\rho_k) \\ \times
\langle \chi_{\alpha vj}(\theta_\alpha;\rho_{k-1})|\langle jM_j |\langle \ell M_\ell | 
\chi_{\alpha'v'j'}(\theta_{\alpha'};\rho_k) \rangle |j' M_j' \rangle | \ell' M_\ell' \rangle
\end{multline}
Unlike the overlap matrix calculated at fixed $\rho$ (\ref{metric}), the overlap matrix of the primitive
functions on the right-hand side of Eq. (\ref{S}) is not symmetric. We denote this matrix as
$\mathsf{O}_\text{SF}(\rho_{k-1},\rho_{k})$. It is discussed in more detail in the next section.

In summary, the adiabatic eigenvalue problem can be solved in three steps. First,
the overlap matrix of the primitive basis functions (\ref{metric}) is constructed and diagonalized
to yield the matrices $\Lambda$ and $\mathsf{X}$. Second, the matrix of the adiabatic Hamiltonian in the primitive basis
is evaluated and transformed to the orthogonalized basis as in Eq. (\ref{MatrixElementsHrho}). Third,
the matrix (\ref{MatrixElementsHrho}) is diagonalized to give the eigenvectors $T_{ni}(\rho)$, which are
convoluted with the sector-to-sector overlap matrix (\ref{S}) to yield the sector-to-sector transformation matrix
$\mathsf{S}(\rho_{k-1},\rho_{k})$.
In order to implement this strategy, we need to evaluate:
(i) the overlap matrix between the primitive functions (\ref{metric}); (ii) the matrix of the adiabatic Hamiltonian
(\ref{MatrixElementsHrho}) in the  primitive basis, and (iii) the
sector-to-sector overlap matrix $\mathsf{O}_\text{SF}(\rho_{k-1},\rho_{k})$ in the primitive basis.

\subsection{Matrix elements in the primitive basis}\label{SIIC}

\subsubsection{Overlap matrix}

Because the primitive functions of the same arrangement are orthonormal, we have
\begin{equation}
\text{O}_{\alpha vjM_j\ell M_\ell,\alpha v'j'M_j' \l'M_\l'}=
\delta_{vv'}\delta_{jj'}\delta_{M_jM_j'}
\delta_{\ell \ell'}\delta_{M_\ell M_\ell'}.
\end{equation}
The only nonzero elements of $\mathsf{O}$ are those between different arrangements.
They have the form
\begin{multline}\label{5D}
\text{O}_{\alpha vjM_j\l M_\l,\alpha' v'j'M_j' \l'M_\l'} =
\int d\hR \int d\hr \int_{0}^{\pi/2}d\theta_\alpha
Y_{jM_j}(\hr) Y_{j'M_j'}(\sr_{\alpha'}) \\ \times
Y_{\l M_\l}(\hR) Y_{\l' M_\l'}(\sR_{\alpha'})
\biggl{(}
\frac{\sin 2\theta_\alpha}{\sin 2\theta_{\alpha'}}
\biggr{)}
 \phi_{\alpha vj}(\theta_\alpha;\rho) 
 \phi_{\alpha' v'j'}(\theta_{\alpha'};\rho),
\end{multline}
where the renormalized FD rovibrational basis functions $\phi_{\alpha vj}(\theta_\alpha;\rho)$ are defined by
\begin{equation}\label{Renormalized}
 \chi_{\alpha vj}(\theta_\alpha;\rho)  =  \frac{2\phi_{\alpha vj}(\theta_\alpha;\rho)}{\sin 2\theta_\alpha}.
\end{equation}
The functions in the integrand (\ref{5D}) depend on different coordinates. A coordinate transformation is therefore
required to express the
functions of one arrangement in terms of the gridpoints of the other. The FD coordinates of different
arrangements are related by kinematic rotations \cite{ParkerPack,Review,Smith}. These rotations
lead to a complicated inseparable dependence of the vectors $\sR_{\alpha'}$, $\sr_{\alpha'}$ on the integration
variables $\hR$, $\hr$. The integral in Eq. (\ref{5D}) is thus a five-dimensional integral, and is computationally
intractable. 

The problem can be simplified by transforming Eq. (\ref{5D}) 
to the BF frame, which allows us to integrate over the three Euler angles analytically.
The transformation of the SF functions has the following form \cite{ParkerPack,Review,Hutson}
\begin{multline}\label{SFBF}
\chi_{\alpha vj}(\theta_\alpha;\rho)Y_{jM_j}(\hr)Y_{\l M_\l}(\hR)
=\chi_{\alpha vj}(\theta_\alpha;\rho)(-)^M(2\l +1)^{1/2} \\ \times  
\sum_{J= |j-\l |}^{j+\l } (2J+1)^{1/2}
\threejm{j}{M_j}{\l}{M_\l}{J}{-M} \sum_{K=-\min(j,J)}^{\min(j,J)} \threejm{j}{K}{\l}{0}{J}{-K} |JMK\rangle |jK\rangle,
\end{multline}
where the BF basis functions are given by
\begin{equation}\label{BFbasis}
\chi_{\alpha vj}(\theta_\alpha;\rho)|JMK\rangle |jK\rangle = \chi_{\alpha vj}(\theta_\alpha;\rho)
\left( \frac{2J+1}{8\pi^2}\right)^{1/2} D^{J*}_{MK}(\Omega_\text{E})
\sqrt{2\pi}Y_{jK}(\gamma_\alpha,0) .
\end{equation}
Here, $K$ is the projection of the total angular momentum $J$ along the BF $z$-axis defined by the vector
$\hR$, the symmetric top functions $|JMK\rangle$ depend on the Euler angles $\Omega_\text{E}$
\cite{ZareBook,Varshalovich}, and the renormalized spherical harmonics $|jK\rangle=\sqrt{2\pi}Y_{jK}(\gamma_\alpha,0)$
describe the rotation of the diatomic molecule in the BF frame \cite{AMW}.

Using Eq. (\ref{SFBF}) and the orthogonality properties of the BF functions,
we can rewrite the SF overlap matrix (\ref{metric}) in the form
\begin{multline}\label{metric2}
\text{O}_{\alpha vjM_j\l M_\l,\alpha' v'j'M_j' \l'M_\l'} = [(2\l +1)(2\l' + 1)]^{1/2}\sum_{J}
(2J +1) \\ \times
\threejm{j}{M_j}{\l}{M_\l}{J}{-M} \threejm{j'}{M_j'}{\l'}{M_\l'}{J}{-M'} 
\sum_{K,K'}(-1)^{K+K'} \threejm{j}{K}{\l}{0}{J}{-K} \threejm{j'}{K'}{\l'}{0}{J}{-K'}  \\ \times 
\langle \chi_{\alpha vj}(\rho;\theta_\alpha) |\langle JMK| \langle jK| 
 \chi_{\alpha' v'j'}(\rho;\theta_{\alpha'})|JMK'\rangle |j'K' \rangle,
\end{multline}
The matrix elements on the right-hand side are much easier to evaluate because the overlap between the symmetric
top functions is $\langle JMK'|JMK\rangle =d^J_{KK'}(\omega_{\alpha\alpha'})$, where the
angle $\omega_{\alpha\alpha'}$ between the vectors $\hR$ and $\sR_{\alpha'}$ is a function of two angles
$\theta_\alpha$ and $\gamma_\alpha$ \cite{Miller,Review}. The result is well known \cite{Review,AMW,Miller}
and we present it here for completeness
\begin{multline}\label{2D}
\langle \chi_{\alpha vj}(\rho;\theta_\alpha) |\langle JMK| \langle jK| 
 \chi_{\alpha' v'j'}(\rho;\theta_{\alpha'})|JMK'\rangle |j'K' \rangle= 2\pi
\int_0^{\pi/2}d\theta_\alpha \int_0^\pi \sin\gamma_\alpha d\gamma_\alpha
\\ \times
\biggl{(}
\frac{\sin 2\theta_\alpha}{\sin 2\theta_{\alpha'}}
\biggr{)}
Y_{jK}(\gamma_\alpha,0)  
\phi_{\alpha vj}(\theta_\alpha;\rho)
d^J_{KK'}(\omega_{\alpha\alpha'})
Y_{j'K'}(\gamma_{\alpha'},0)
\phi_{\alpha' v'j'}(\theta_{\alpha'};\rho)
\end{multline}

\subsubsection{Adiabatic Hamiltonian and sector-to-sector overlap matrices}

The matrix elements of the adiabatic Hamiltonian (\ref{Had}) can be derived following the same procedure.
The field-independent part of the adiabatic Hamiltonian (\ref{Had}) can be written in the BF frame as \cite{Hutson}
\begin{equation}\label{HBF}
\tilde{H}_\text{fi}(\rho)= \frac{ (\bf{J} - \bf{j}_\alpha)^2 }{2\mu\rho^2\cos^2\theta_\alpha}
+V(\rho,\,\theta_\alpha,\,\gamma_\alpha) - V_{\alpha}(r_{\alpha}) +\tilde{H}_\text{mol},
\end{equation}
This Hamiltonian is formally equivalent to that given by (\ref{Had}). Therefore, we can evaluate
the matrix elements of $H_\text{fi}(\rho)$ in the SF frame using the transformation (\ref{SFBF}) as follows
\begin{multline}\label{potential2}
\langle \chi_{\alpha vj}(\theta_\alpha;\rho)|\langle jM_j |\langle \l M_\l | \tilde{H}_\text{fi}(\rho)|
\chi_{\alpha'v'j'}(\theta_{\alpha'};\rho) \rangle |j' M_j' \rangle | \l' M_\l'\rangle = [(2\l +1)(2\l' + 1)]^{1/2}\sum_{J}
(2J +1) \\ \times
\threejm{j}{M_j}{\l}{M_\l}{J}{-M} \threejm{j'}{M_j'}{\l'}{M_\l'}{J}{-M'} 
\sum_{K,K'}(-1)^{K+K'} \threejm{j}{K}{\l}{0}{J}{-K} \threejm{j'}{K'}{\l'}{0}{J}{-K'} \\ \times
\langle \chi_{\alpha vj}(\rho;\theta_\alpha) |\langle JMK| \langle jK| \tilde{H}_\text{fi}(\rho) |
 \chi_{\alpha' v'j'}(\rho;\theta_{\alpha'})|JMK'\rangle |j'K' \rangle,
\end{multline}
Since the interaction potential does not depend on the Euler angles, the matrix elements of the field-independent
part of the adiabatic Hamiltonian in the BF basis can be reduced to 2D integrals over $\theta_\alpha$ and $\gamma_\alpha$.
They have the form \cite{ABC,Review,AMW}
\begin{multline}\label{potentialBF}
\langle \chi_{\alpha vj}(\rho;\theta_\alpha) |\langle JMK| \langle jK| \tilde{H}_\text{fi}(\rho) |
 \chi_{\alpha' v'j'}(\rho;\theta_{\alpha'})|JMK'\rangle |j'K' \rangle= 2\pi
\int_0^{\pi/2}d\theta_\alpha \int_0^\pi \sin\gamma_\alpha d\gamma_\alpha
\\ \times
Y_{jK}(\gamma_\alpha,0)  
\phi_{\alpha vj}(\theta_\alpha;\rho)
d^J_{KK'}(\omega_{\alpha\alpha'}) \left[ \frac{ ({\bf J} - {\bf j}_\alpha)^2 }{2\mu\rho^2\cos^2\theta_\alpha}
+V(\rho,\,\theta_\alpha,\,\gamma_\alpha) - V_{\alpha}(r_{\alpha}) + \tilde{H}_\text{mol} \right] \\ \times
\biggl{(}
\frac{\sin 2\theta_\alpha}{\sin 2\theta_{\alpha'}}
\biggr{)}
Y_{j'K'}(\gamma_{\alpha'},0) \phi_{\alpha' v'j'}(\theta_{\alpha'};\rho).
\end{multline}
The matrix elements of the first term in square brackets can be obtained by expressing the angular momentum operator
 in terms of the raising and lowering operators, and taking into account the anomalous commutation relations
as described in Refs. \cite{ParkerPack,Review,ZareBook,Miller}. The remaining terms in Eq. (\ref{potentialBF})
are very similar to the overlap matrix elements discussed in the previous section.

Finally, the sector-to-sector overlap matrix in the primitive basis has the form
\begin{multline}\label{Os}
\langle \chi_{\alpha vj}(\theta_\alpha;\rho_{k-1})|\langle jM_j |\langle \l M_\l | 
\chi_{\alpha'v'j'}(\theta_{\alpha'};\rho_k) \rangle |j' M_j' \rangle | \l' M_\l'\rangle =
[(2\l +1)(2\l' + 1)]^{1/2}\sum_{J} (2J +1) \\ \times
\threejm{j}{M_j}{\l}{M_\l}{J}{-M} \threejm{j'}{M_j'}{\l'}{M_\l'}{J}{-M'} 
\sum_{K,K'}(-1)^{K+K'} \threejm{j}{K}{\l}{0}{J}{-K} \threejm{j'}{K'}{\l'}{0}{J}{-K'} \\ \times
\langle \chi_{\alpha vj}(\rho_{k-1};\theta_\alpha) |\langle JMK| \langle jK |
 \chi_{\alpha' v'j'}(\rho_k;\theta_{\alpha'})|JMK'\rangle |j'K' \rangle,
\end{multline}
where the BF sector-to-sector overlap matrix is given by
\begin{multline}\label{OsBF}
\langle \chi_{\alpha vj}(\rho_{k-1};\theta_\alpha) |\langle JMK| \langle jK| 
 \chi_{\alpha' v'j'}(\rho_k;\theta_{\alpha'})|JMK'\rangle |j'K' \rangle= 2\pi
\int_0^{\pi/2}d\theta_\alpha \int_0^\pi \sin\gamma_\alpha d\gamma_\alpha
\\ \times
\biggl{(}
\frac{\sin 2\theta_\alpha}{\sin 2\theta_{\alpha'}}
\biggr{)}
Y_{jK}(\gamma_\alpha,0)  
\phi_{\alpha vj}(\theta_\alpha;\rho_{k-1})
d^J_{KK'}(\omega_{\alpha\alpha'}) 
Y_{j'K'}(\gamma_{\alpha'},0) \phi_{\alpha' v'j'}(\theta_{\alpha'};\rho_k).
\end{multline}

\subsection{Interaction with electric fields}

If the interaction of the three-body term with electric fields is neglected, Eq. (\ref{Hef})
can be rewritten as
\begin{equation}\label{HefJacobi}
\tilde{H}_{\text{ef}} = -d_\alpha (\rho,\theta_\alpha) E\cos\chi_\alpha,
\end{equation}
where $\chi_\alpha$ is the polar SF angle of the diatomic molecule in arrangement $\alpha$. In the limit
of large $\rho$, there is no overlap between the vibrational basis functions of different arrangements, and
the matrix elements of the interaction with electric fields (\ref{HefJacobi}) can be written as
\begin{multline}\label{MatrixE1}
\langle \chi_{\alpha vj}(\theta_\alpha;\rho)|\langle jM_j |\langle \l M_\l | \tilde{H}_{\text{ef}}|
\chi_{\alpha'v'j'}(\theta_{\alpha'};\rho) \rangle |j' M_j' \rangle | \l' M_\l'\rangle 
= -E \delta_{\alpha \alpha'}\delta_{\l \l'}\delta_{M_\l M_\l'} \\ \times
\langle \chi_{\alpha vj}(\theta_\alpha;\rho)|d_\alpha(\rho,\theta_\alpha)| \chi_{\alpha v'j'}(\theta_\alpha;\rho)\rangle
\langle jM_j | \cos\chi_{\alpha} |  j'M_j'\rangle.
\end{multline}
Furthermore, at large $\rho$ the dipole moment function $d_\alpha(\theta_\alpha;\rho)$ is localized near the equilibrium distance
of the diatomic molecule and it is a slowly varying function of $r_\alpha$. Therefore, it is a good approximation to
neglect the matrix elements off diagonal in $v$:
\begin{equation}\label{MatrixE1a}
\langle \chi_{\alpha vj}(\theta_\alpha;\rho)|d_\alpha(\rho,\theta_\alpha)| \chi_{\alpha v'j'}(\theta_\alpha;\rho)\rangle =
\delta_{vv'}d_{\alpha v}.
\end{equation}
We note that this approximation is consistent with neglecting the coupling between different arrangements due to
the electric field (\ref{MatrixE1}).
For low $v$ considered in this work, the vibrational dependence of the matrix element in Eq.~(\ref{MatrixE1a})
is very weak. For example, the dipole moment of LiF in the $v=8$ vibrational state is only 11\% larger than
for the ground vibrational state \cite{DipoleMoments}. We therefore assume that the matrix
element in Eq. (\ref{MatrixE1a}) is equal to the dipole moment of the molecule in the ground vibrational
state and is independent of $v$.
Writing $\cos\chi_\alpha = (4\pi/3)^{1/2}Y_{10}(\hr)$ and evaluating the integral over the product of three spherical
harmonics \cite{ZareBook,Varshalovich}, we find \cite{OurWork,Volpi}
\begin{multline}\label{MatrixE2}
\langle \chi_{\alpha vj}(\theta_\alpha;\rho)|\langle jM_j |\langle \l M_\l | \tilde{H}_{\text{ef}}|
\chi_{\alpha'v'j'}(\theta_{\alpha'};\rho) \rangle |j' M_j' \rangle | \l' M_\l'\rangle 
 = - E d_\alpha \delta_{\alpha\alpha'}\delta_{vv'}\delta_{\l \l'}\delta_{M_\l M_\l'} \\ \times
(-)^{M_j}[(2j+1)(2j'+1)]^{1/2} \threejm{j}{-M_j}{1}{0}{j'}{M_j'}  \threejm{j}{0}{1}{0}{j'}{0},
\end{multline}
where the symbols in parentheses are the 3$-j$ symbols. In order to complete the definition of the matrix elements
of the adiabatic Hamiltonian (\ref{Had}), we need to evaluate the matrix elements of the asymptotic Hamiltonian (\ref{Hmol}).
Because the overlap matrix (\ref{metric2}) becomes diagonal in the asymptotic region, the matrix elements of the asymptotic
Hamiltonian can be evaluated using Eq. (\ref{MatrixE2}), 
\begin{multline}\label{MatrixHas}
\langle \chi_{\alpha vj}(\theta_\alpha;\rho)|\langle jM_j |\langle \l M_\l | \tilde{H}_{\text{as}}|
\chi_{\alpha'v'j'}(\theta_{\alpha'};\rho) \rangle |j' M_j' \rangle | \l' M_\l'\rangle 
  = \delta_{\alpha\alpha'}\delta_{vv'}\delta_{\l \l'}\delta_{M_\l M_\l'} \\ \times
 \left[ \delta_{jj'}\delta_{M_jM_j'}\epsilon_{\alpha vj}(\rho) -
Ed_\alpha  (-)^{M_j} [(2j+1)(2j'+1)]^{1/2}\threejm{j}{-M_j}{1}{0}{j'}{M_j'}  \threejm{j}{0}{1}{0}{j'}{0} \right]
\end{multline}
This expression shows that the asymptotic Hamiltonian is not diagonal in the uncoupled
SF basis (\ref{primitive}), except for $E = 0$. The scattering boundary conditions must be applied
in the basis that diagonalizes the asymptotic Hamiltonian (\ref{MatrixHas}) \cite{Roman2004}. The eigenfunctions of $\tilde{H}_\text{as}$
\begin{equation}\label{AsymptoticE}
\tilde{H}_\text{as}\psi_{\alpha v\tau M_\tau}(\hr, \theta_\alpha;\rho) = \epsilon_{\alpha v\tau M_\tau}(\rho)
\psi_{\alpha v\tau M_\tau}(\hr, \theta_\alpha;\rho)
\end{equation}
can be written as linear combinations of the basis functions given by Eq. (\ref{primitive})
\begin{equation}\label{AsymptoticFunctions}
\psi_{\alpha v\tau M_\tau}(\hr, \theta_\alpha;\rho) =
\sum_{j,M_j}C_{jM_j, \tau M_\tau}(E)\chi_{\alpha v j}(\theta_\alpha;\rho) Y_{jM_j}(\hr),
\end{equation}
where the field-dependent mixing coefficients $C_{jM_j, \tau M_\tau}(E)$ can be obtained by numerical
diagonalization of the asymptotic Hamiltonian matrix (\ref{MatrixHas}). Since the electric field only couples 
the levels with $\Delta j=\pm 1$, the dependence on $j$ of the overlap of rovibrational FD basis functions 
in Eq. (\ref{MatrixHas}) can be neglected, and Eq. (\ref{AsymptoticFunctions}) rewritten as
\begin{equation}\label{AF}
\psi_{\alpha v\tau M_\tau}(\hr, \theta_\alpha;\rho) = \chi_{\alpha v \tau}(\theta_\alpha;\rho)
\sum_{j,M_j}C_{jM_j, \tau M_\tau}(E) Y_{jM_j}(\hr),
\end{equation}
where $\tau,M_\tau$ labels the field-dressed states. The index $\tau$ of the function $\chi_{\alpha v \tau}(\theta_\alpha;\rho)$
denotes the dominant rotational state in the expansion (\ref{AF}).

The matrix of the asymptotic Hamiltonian (\ref{HmolJacobi}) can also be written in
the Jacobi basis defined in the same way as the FD basis (\ref{primitive}) 
\begin{equation}\label{primitiveJacobi}
\frac{\xi_{\alpha vj}(r_\alpha)}{r_\alpha}|jM_j\rangle |{\l M_\l}\rangle,
\end{equation}
where the functions $|jM_j\rangle$ and $|\ell M_\ell\rangle$ are the same as in Eq. (\ref{primitive}) and
the renormalized Jacobi rovibrational basis functions are the radial
eigenfunctions of the diatomic molecule (\ref{HmolJacobi1})
in the absence of an electric field
\begin{equation}\label{Xi}
\biggl{[}
-\frac{1}{2\mu}\frac{\partial^2}{\partial r_\alpha^2} + \frac{{ j(j+1)}}{2\mu r_\alpha^2}+V(r_\alpha)
\biggr{]}\xi_{\alpha vj} (r_\alpha)= \epsilon_{\alpha vj}\xi_{\alpha vj} (r_\alpha).
\end{equation}
The eigenfunctions of $H_\text{mol}$ in Eq. (\ref{HmolJacobi}) are given by $\xi_{\alpha vj} (r_\alpha)/r_\alpha$.

Although the basis (\ref{primitiveJacobi}) is slowly converging at small $\rho$ (where different reaction
arrangements are strongly coupled \cite{Miller,ParkerPack,Review}), it is convenient in the asymptotic region,
where the Jacobi basis functions of different arrangements are orthonormal.
The matrix representation of the asymptotic Hamiltonian (\ref{Hef}) in this basis is given by
\begin{multline}\label{MatrixHasJacobi}
\langle \xi_{\alpha vj}(r_\alpha) | \langle jM_j | \langle \l M_\l | H_{\text{as}}|
\xi_{\alpha' v'j'}(r_{\alpha'}) \rangle | j'M_j'\rangle | \l' M_\l' \rangle
 = \delta_{\alpha\alpha'}\delta_{vv'}\delta_{\l \l'}\delta_{M_\l M_\l'} \\ \times
 \left[ \delta_{jj'}\delta_{M_jM_j'}\epsilon_{\alpha vj} -
Ed_\alpha (-)^{M_j} [(2j+1)(2j'+1)]^{1/2} \threejm{j}{-M_j}{1}{0}{j'}{M_j'}  \threejm{j}{0}{1}{0}{j'}{0} \right],
\end{multline}
where $\epsilon_{\alpha vj}$ is the rovibrational energy of the diatomic molecule defined by Eq. (\ref{Xi}).

Since $\epsilon_{\alpha vj}(\rho)\to \epsilon_{\alpha vj}$ in the limit of large $\rho$, Eqs. (\ref{MatrixHasJacobi})
and (\ref{MatrixHas}) define {\it the same} matrix, and the eigenvectors $C_{jM_j,\tau M_\tau}$ in Eq. (\ref{AF})
are identical in the Jacobi and FD coordinates. This is important for reactive
scattering boundary conditions (see Sec. \ref{SIIF}). 

\subsection{Propagation}\label{SIIE}

In order to determine the expansion coefficients $F_i(\rho)$, it is necessary to integrate
the CC equations (\ref{CC}) from small values of $\rho$ to the asymptotic region. Since the adiabatic basis
(\ref{AEP}) changes with $\rho$, the solutions $F_i(\rho)$ should be transformed to a new basis as $\rho$ increases.
The integration interval of $\rho$ is usually divided into small sectors such that the adiabatic
basis (\ref{AEP}) does not change within the sector. At the boundary between the $(k-1)$-th and $k$-th sectors,
the solutions of Eq. (\ref{CC}) must be transformed to the basis of the $k$-th sector.
The transformation is determined by the requirement that the total wave function (\ref{TotalWF})
be continuous at the boundary. The coefficients $F_i(\rho_k)$ can be written as
\begin{equation}\label{ActionS}
F_i(\rho_k) = \sum_{i'}F_{i'}(\rho_{k-1}) [\mathsf{S}^\text{T}(\rho_{k-1},\rho_{k})]_{ii'},
\end{equation}
where $\rho_{k-1}$ and $\rho_{k}$ denote the centers of the respective sectors, and
$\mathsf{S}(\rho_{k-1},\rho_{k})$ is the sector-to-sector overlap matrix given by Eq. (\ref{Sdefinition}).

To preserve the numerical stability of the solutions of the CC equations, we propagate the log-derivative matrix
\cite{Johnson,Manolopoulos,ParkerPack,Review} defined as
\begin{equation}\label{Y}
\mathsf{Y}(\rho_{k}) = \mathsf{F}'(\rho_{k}) [\mathsf{F}(\rho_k)]^{-1},
\end{equation}
where $\mathsf{F}(\rho)$ is the matrix with {\it columns} represented by the solutions to Eq. (\ref{CC}),
and the prime indicates differentiation with respect to $\rho$.
We propagate the matrix $\mathsf{Y}$ across the sector $[\rho_{k-2},\rho_{k-1}]$
(the first sector of the grid is $[\rho_0,\rho_1]$) using the diagonal
reference potential propagator of Manolopoulos \cite{Manolopoulos}. After reaching the
right end at $\rho_{k-1}$, we transform the log-derivative matrix to the adiabatic basis of
the next sector $[\rho_{k-1},\rho_k]$.  Eq. (\ref{ActionS}) can be written in matrix form as
\begin{equation}
\mathsf{F}(\rho_k) = \mathsf{S}^\text{T}(\rho_{k-1},\rho_{k}) \mathsf{F}(\rho_{k-1}).
\end{equation}
Using this relation and the definition (\ref{Y}), we obtain
\begin{equation}\label{SY}
\mathsf{Y}(\rho_k) = \mathsf{S}^\text{T}(\rho_{k-1},\rho_k) \mathsf{Y}(\rho_{k-1})
\mathsf{S}(\rho_{k-1},\rho_k).
\end{equation}
The CC equations can thus be integrated by repeated application of the log-derivative
propagation \cite{Manolopoulos} and the transformation (\ref{SY}).

\subsection{Asymptotic Boundary Conditions}\label{SIIF}

In the asymptotic limit $\rho\to\infty$, different reaction arrangements become uncoupled, and the reactive
scattering wave function can be conveniently re-expressed in the SF Jacobi basis
\begin{equation}\label{SpaceFixedJacobi}
\Psi = \sum_{\alpha, v,\tau,M_\tau}\sum_{\l, M_\l} \frac{1}{r_\alpha R_\alpha}F_{\alpha v\tau M_\tau \l M_\l}(R_\alpha)
\psi_{\alpha v \tau M_\tau }({\bf r}_\alpha) Y_{\l M_\l}(\hR),
\end{equation}
where the asymptotic field-dressed Jacobi functions of the diatomic molecule are given by 
\begin{equation}\label{A0}
H_\text{as}\psi_{\alpha v\tau M_\tau}({\bf r}_\alpha) = \epsilon_{\alpha v\tau M_\tau}\psi_{\alpha v\tau M_\tau}({\bf r}_\alpha).
\end{equation}
The energies $\epsilon_{\alpha v\tau M_\tau}$ are the Stark levels of the diatomic molecule in arrangement $\alpha$.
The field-dressed eigenfunctions in the Jacobi coordinates defined by Eq. (\ref{A0}) are distinct from the functions
in the FD coodinates
(\ref{AsymptoticE}), and we will specify the arguments of both functions to avoid confusion. Similarly to Eq. (\ref{AsymptoticFunctions}),
the Jacobi asymptotic functions can be expanded in spherical harmonics
\begin{equation}\label{A1}
\psi_{\alpha v\tau M_\tau}({\bf r}_\alpha) = \xi_{\alpha v \tau}(r_\alpha) \sum_{j,M_j}C_{jM_j, \tau M_\tau}(E)Y_{jM_j}(\hr)
\end{equation}
with the expansion coefficients $C_{jM_j, \tau M_\tau}(E)$ given by Eq. ({\ref{AF}). In Eq. (\ref{A1}),
$\xi_{\alpha v j}(r_\alpha)$ is the Jacobi ro-vibrational eigenfunction in arrangement $\alpha$ given by Eq. (\ref{Xi}).
The radial functions in Eq. (\ref{SpaceFixedJacobi}) have the following asymptotic behavior \cite{ParkerPack,Lester}
\begin{align}\label{F}\notag
F^{\alpha v\tau M_\tau \l M_\l}_{\alpha' v'\tau' M_\tau' \l' M_\l'}(R_{\alpha'}\to\infty) &\simeq 
\delta_{\alpha\alpha'}\delta_{vv'}\delta_{\tau\tau'}\delta_{M_\tau M_\tau'}\delta_{\l\l'}\delta_{M_\l M_\l'}
\exp[-\text{i}(k_{\alpha v\tau M_\tau}R_\alpha - \l \pi/2)] \\ 
&- \left( \frac{k_{\alpha v\tau M_\tau}}{k_{\alpha'v'\tau'M_\tau'}} \right)^{1/2}
S_{\alpha'v'\tau' M_\tau' \l' M_\l';\alpha v\tau M_\tau \l M_\l}\exp[\text{i}(k_{\alpha' v'\tau'M_\tau'}R_{\alpha'} - \l' \pi/2)]
\end{align}
where $k_{\alpha v \tau M_\tau}^2=2\mu (E_\text{t} - \epsilon_{\alpha v\tau M_\tau})$ is the asymptotic wave vector given in
terms of the total energy $E_\text{t}$ and the asymptotic Stark energy (\ref{A0}), and
 $S_{ij}$ are the $S$-matrix elements. They can be evaluated by matching the asymptotic form of the FD wave function
(\ref{TotalWF}) to the Jacobi wave function (\ref{SpaceFixedJacobi}). The details of the asymptotic matching procedure
are described in the Appendix.

The scattered part of the wave function (\ref{SpaceFixedJacobi}) corresponding to the initial flux of
molecules in the state ($\alpha, v,\tau, \l$) propagating along the direction $\hat{R}_{\alpha_\text{i}}$
has the form \cite{ParkerPack, Review,Lester}
\begin{multline}\label{ScatteringAmplitude}
\Psi_{\alpha v\tau M_\tau}^\text{scattered} (R_{\alpha'} \to \infty)\simeq \sum_{\alpha', v',\tau',M_\tau'} \text{i}
(k_{\alpha v\tau M_\tau}k_{\alpha'v'\tau'M_\tau'})^{-1/2}
q_{\alpha v\tau M_\tau \to \alpha' v'\tau' M_\tau'}(\hat{R}_{\alpha_\text{i}},\sR_{\alpha'}) \\ \times
\frac{\exp[\text{i}k_{\alpha'v'\tau'M_\tau'}R_{\alpha'}] }{r_{\alpha'} R_{\alpha'}}\psi_{\alpha'v'\tau'M_\tau'}({\bf r}_\alpha),
\end{multline}
where $q_{\alpha v\tau M_\tau \to \alpha' v'\tau' M_\tau'}(\hat{R}_{\alpha_\text{i}},\sR_{\alpha'})$ is the
the scattering amplitude. An expression for the scattering amplitude in terms of the $S$-matrix elements
can be obtained by substituting Eq. (\ref{F}) into Eq. (\ref{SpaceFixedJacobi}). After separating the
scattered component of the wave function and comparing the result with Eq. (\ref{ScatteringAmplitude}),
we find \cite{Roman2004}
\begin{multline}\label{ScatteringAmplitude2}
q_{\alpha v\tau M_\tau \to \alpha' v'\tau' M_\tau'}(\hat{R}_{\alpha_\text{i}},\sR_{\alpha'})
=2\pi \sum_{\l, M_\l}\sum_{\l', M_\l'} \text{i}^{\l-\l'} Y_{\l M_\l}^*(\hat{R}_{\alpha_\text{i}})
Y_{\l M_\l}(\sR_{\alpha'}) \\ \times
\left[ \delta_{\alpha\alpha'}\delta_{vv'}\delta_{\tau\tau'}\delta_{M_\tau M_\tau'}
\delta_{\l \l'} \delta_{M_\l M_\l'} - S_{\alpha v\tau M_\tau \l M_\l; \alpha' v'\tau' M_\tau'\l' M_\l'} \right].
\end{multline}
The differential cross section for reactive transitions between the field-dressed states is given by
\begin{equation}\label{DCSreactive}
\frac{d\sigma_{\alpha v\tau M_\tau\to \alpha' v'\tau' M_\tau'}}{d\hat{R}_{\alpha_\text{i}}d\sR_{\alpha'}}
= \frac{1}{k_{\alpha v\tau M_\tau}^2}
|q_{\alpha v\tau M_\tau \to \alpha' v'\tau' M_\tau'}(\hat{R}_{\alpha_\text{i}},\sR_{\alpha'})|^2.
\end{equation}
Integrating this expression over $\sR_{\alpha'}$ and averaging over $\hat{R}_{\alpha_\text{i}}$ gives
the integral reaction cross section
\begin{equation}\label{ICS}
\sigma_{\alpha v\tau M_\tau\to \alpha' v'\tau' M_\tau'}
= \frac{\pi}{k_{\alpha v\tau M_\tau}^2} \sum_{\l,M_\l}\sum_{\l',M_\l'}
| \delta_{\alpha\alpha'}\delta_{vv'}\delta_{\tau\tau'}\delta_{M_\tau M_\tau'}
\delta_{\l \l'} \delta_{M_\l M_\l'} - S_{\alpha v\tau M_\tau \l M_\l; \alpha' v'\tau' M_\tau'\l' M_\l'} |^2.
\end{equation}
Taking into account the conservation of the total angular momentum projection $M$, we can rewrite this
expression as a sum of partial cross sections calculated from the $S$-matrix elements at fixed $M$
\begin{equation}\label{ICSa}
\sigma_{\alpha v\tau M_\tau\to \alpha' v'\tau' M_\tau'}
= \frac{\pi}{k_{\alpha v\tau M_\tau}^2}\sum_M\sum_{\l,M_\l}\sum_{\l',M_\l'}
| \delta_{\alpha\alpha'}\delta_{vv'}\delta_{\tau\tau'}\delta_{M_\tau M_\tau'}
\delta_{\l \l'} \delta_{M_\l M_\l'} - S^M_{\alpha v\tau M_\tau \l M_\l; \alpha' v'\tau' M_\tau'\l' M_\l'} |^2.
\end{equation}
where the summation over $M_\ell$ and $M_\ell'$ is restricted so that $M_\ell + M_\tau = M_\ell' + M_\tau' = M$.
Eqs. (\ref{ScatteringAmplitude2})--(\ref{ICSa}) generalize the expressions derived by Krems and Dalgarno
\cite{Roman2004} to reactive scattering in external fields.
In the absence of an external magnetic field, the Stark levels with energies $\epsilon_{\alpha v\tau M_\tau}$ and $\epsilon_{\alpha v\tau -M_\tau}$ 
are degenerate. Because the sum $j+j'+1$ in Eq. (\ref{MatrixHas}) is always even, the matrix elements of the asymptotic Hamiltonian 
do not depend on the sign of $M$ and so do the $M$-resolved cross sections given by Eq. (\ref{ICSa}). With this in mind,
Eq. (\ref{ICSa}) can be rewritten as 
\begin{multline}\label{ICS2}
\sigma_{\alpha v\tau M_\tau\to \alpha' v'\tau' M_\tau'}
= \frac{\pi}{k_{\alpha v\tau M_\tau}^2} \sum_{M\ge 0} \sum_{\l,M_\l}\sum_{\l',M_\l'} (2-\delta_{M,0}) \\ \times
|\delta_{\alpha\alpha'}\delta_{vv'}\delta_{\tau\tau'}\delta_{M_\tau M_\tau'}
\delta_{\l \l'} \delta_{M_\l M_\l'} - S^M_{\alpha v\tau M_\tau \l M_\l; \alpha' v'\tau' M_\tau'\l' M_\l'}|^2
\end{multline}

\section{Results and discussion}

In this section, we present the results of preliminary calculations based on the theory described in Sec. II. We
study the variation of the cross sections for the chemical reactions LiF($v=1,j=0$)~+~H $\to$ LiF + H and
Li~+~HF($v=0,j=0$) $\to$ LiF + H with the collision energy and the strength of an applied electric field. We also
analyze the competing process of vibrational relaxation in LiF ($v=1,j=0$)+ H collisions
and suggest a mechanism for electric field control of chemical reactions at low temperatures.

\subsection{Computational details}

The hyperspherical basis functions $\chi_{\alpha vj}(\theta_\alpha;\rho)$ are constructed by solving Eq. (\ref{chis})
using the Fourier grid Hamiltonian method \cite{ColbertMiller}. The functions become more localized with
increasing $\rho$. The localization region of the functions (\ref{chis}) can be defined as follows \cite{ABC,Private}
\begin{equation}\label{theta_endpoints}
\theta_\alpha^\text{min} = \sin^{-1}\left[{r_\alpha^\text{min}}/{\rho}\right];\quad
\theta_\alpha^\text{max} = \sin^{-1}\left[{r_\alpha^\text{max}}/{\rho}\right]. \\
\end{equation}
We chose the endpoints $r_\alpha^\text{min}$ and $r_\alpha^\text{max}$
so that the rovibrational function of the
diatomic molecule (\ref{Xi}) is nonzero in the interval $[r_\alpha^\text{min},r_\alpha^\text{max}]$.
This guarantees that the functions $\chi_{\alpha vj}(\theta_\alpha;\rho)$ vanish outside the
interval $[\theta_\alpha^\text{min},\theta_\alpha^\text{max}]$, and the integration range
in Eqs. (\ref{2D}) and (\ref{potentialBF}) reduces from $[0,\pi/2]$ to $[\theta_\alpha^\text{min},\theta_\alpha^\text{max}]$.
In addition, using the $\rho$-dependent endpoints (\ref{theta_endpoints}) ensures the accuracy at large $\rho$ which is necessary
to calculate reactive scattering cross sections at low temperatures.

The BF matrix elements given by Eqs. (\ref{2D}), (\ref{potentialBF}) and (\ref{OsBF}) are evaluated at fixed $J$ as described
by Miller \cite{Miller}, and Alexander, Manolopoulos and Werner \cite{AMW}.
The integration is performed with 45 Gauss-Legendre quadrature points
in $\theta_\alpha \in [\theta_\alpha^\text{min}(\rho),\theta_\alpha^\text{max}(\rho)]$ and $\gamma_\alpha \in [0,\pi]$.
The basis functions are transformed between different reaction arrangements using the expressions derived in Refs. \cite{Review,Miller}.
The upper limit of the total angular momentum $J_\text{max}=j_\text{max}+\l_\text{max}$ is determined
by the BF to SF transformation (\ref{SFBF}). The overlap and Hamiltonian matrices are stored on the hard disk for subsequent
transformation to the SF frame. At large hyperradius, the couplings between different arrangements become negligibly small, and
the Hamiltonian and sector-to-sector overlap matrices become sparse. We therefore use sparse matrix storage and retain
all matrix elements with the absolute magnitude larger than $10^{-4}$ cm$^{-1}$ at $\rho >6.6$ $a_0$.
At each propagation step, the field-independent Hamiltonian and overlap matrices are constructed
from the BF matrices using Eqs. (\ref{potential2}) and (\ref{Os}). The matrix elements of the interaction with electric fields
are computed directly in the SF basis using Eqs. (\ref{MatrixE2}) and added to the field-independent Hamiltonian.

In order to avoid the overcompleteness problem \cite{SchatzKuppermann} at small hyperradius, the eigenvalues
of the overlap matrix smaller than the tolerance parameter $\varepsilon=0.1$ are discarded. The remaining
eigenvectors of the overlap matrix 
are used to transform the Hamiltonian matrix from the primitive to orthogonalized basis (\ref{potential2}) as described above
[see Sec. II B, Eq. (\ref{Orthogonalization})]. The eigenvalue problem (\ref{MatrixElementsHrho}) is solved for the
transformed Hamiltonian
and the $T_{ni}(\rho)$ coefficients are used to assemble the sector-to-sector transformation matrix via Eq.~(\ref{S}), which
is used to transform the log-derivative matrix to the next propagation step (\ref{SY}).
At the end of the propagation, the log-derivative matrix is transformed back to the SF primitive basis (\ref{primitive}) and then
to the asymptotic basis (i.e. the representation in which the asymptotic Hamiltonian is diagonal \cite{Roman2004}).
After these transformations, the wave function of the reaction complex is matched to the asymptotic
Jacobi functions to yield the reactance $K$-matrix as described in the Appendix. The cross sections and probabilities for
the reaction are obtained from the $K$ and $S$-matrices using the expressions derived in Sec. II F.

For the LiHF system we employed the most recent potential energy surface of Aguado, Paniagua  and Werner \cite{Werner},
and used the dipole moments 6.33 D for LiF and 1.83 D for HF from Refs. \cite{DipoleMoments}. Our largest primitive basis (\ref{primitive}) 
included the vibrational states of LiF and HF with $v_\text{LiF}\le 8$ and $v_\text{HF}\le 3$, augmented with 8 rotational states ($j\le 7$) and 5 partial waves ($\ell\le 4$), which resulted in 1920 coupled channels for $M=0$. The number of scattering channels is much larger
in the SF basis, and in order to make the reactive scattering calculations feasible, we had to reduce the basis set parameters
recommended by Weck and Balakrishnan \cite{BalaLiF} for $J=0$ calculations.
We performed the calculations in a cycle over the total angular momentum projection in the range $M=0-4$.

In order to test our program, we computed the cross sections for the Li + HF reaction at zero
electric field with two different methods: (i) the SF uncoupled formalism described in this paper, and
(ii) the standard method based on the total angular momentum representation in the BF frame \cite{SchatzKuppermann,Review,ABC,David}.
The cross sections computed with method (ii) were compared with the calculations using the ABC code of
Manolopoulos and co-workers \cite{ABC}, and  a good agreement was found for the total reaction probabilities
at all $J$. The cross sections obtained with method (i)
were summed over $\ell$ and the cross sections obtained with method (ii) were summed over $J$. 
In both calculations, we used the maximum number of rotational states $j_\text{max}=2$ and included
6 partial waves ($\ell=0-5$) in basis (i) and 5 total angular momenta ($J=0-5$) in basis (ii). The propagation parameters
were (in units of $a_0$): $\rho_\text{min}=3.4$, $\rho_\text{max}=30.5$, and $\Delta \rho = 0.02$.
Figure 1 shows that the total reaction cross sections calculated with methods (i) and (ii)
are in good agreement, which demonstrates that both the SF and BF formalisms are implemented correctly. 
Note that the SF basis is restricted to $\ell_\text{max}=5$, whereas the BF cross section for $J=4$
contains contributions from $\ell=6$. This leads to a small discrepancy between the SF and BF results at a collision energy
of $\sim 0.2$ K where the $J=4$ contribution is most significant.

\subsection{Numerical results}

Electric fields couple different rotational states of the reactants and products and lead to the formation of the
field-dressed pendular states \cite{Bretislav}. The strength of the molecule-field coupling can be quantified as $Ed/B_e$,
where $B_e$ is the rotational constant of the molecule. The rotational constant of HF (20.96 cm$^{-1}$) is large compared to that of LiF (1.35 cm$^{-1}$),
and the dipole moment of HF (1.83 D) is significantly smaller than that of LiF (6.33 D). Therefore, the molecule-field coupling is much
stronger in the entrance channel of the LiF + H $\to$ Li~+~HF reaction. Figure 2 shows that in an electric field of 200 kV/cm,
the Stark shift of the $v=1,j=0$ level of LiF amounts to 14.5 cm$^{-1}$, whereas the ground ro-vibrational state of HF is shifted
only by 0.3 cm$^{-1}$.

We consider the reaction of LiF molecules in the lowest-energy Stark state, which correlates to the
state $|v=1,j=0\rangle$ in the zero-field limit. The energy of this state decreases with increasing the field.
As the rotational levels of the HF product are only slightly modified by the electric field (Fig. 2),
the exothermicity of the LiF($v=0,j=0$) + H reaction forming HF$(v'=0,j'=0)$ molecules decreases from
240.57 cm$^{-1}$ at zero field to 226.39 cm$^{-1}$ at $E=200$ kV/cm. This example shows that the exothermicity of
state-resolved
chemical reactions can be controlled with electric fields. If the energy defect between the initial and final ro-vibrational
levels is small, it may be possible to close or open a particular reaction channel by varying the electric field
strength. For example, the reactive channel $|v=1,\tau=2,M_\tau=2\rangle \to |v'=0,\tau'=3\rangle$ shown in Fig. 2
becomes closed as the field increases from zero to $E=150$ kV/cm, and reopens at $E>200$ kV/cm (the reader is reminded that
the field-dressed state $|\tau M_\tau\rangle$ corresponds to the field-free rotational state $|jM_j\rangle$ in the limit of zero
electric field).
As in the case of photodissociation \cite{SO2} and predissociation \cite{RomanPRL2004}, this suggests that
reactive scattering cross sections near threshold can be efficiently manipulated with electric fields.

Figure 3 (upper panel) shows the total reaction cross section as a function of the collision energy.
The cross section at zero field is small due to the suppression of tunneling of the heavy F atom under the reaction
barrier \cite{BalaReview,BalaLiF}. Figure 3 illustrates two important observations. First, electric fields enhance the reaction probability
by several orders of magnitude over a large interval of collision energies. The largest effect is observed in the $s$-wave scattering
regime, where the electric field modifies the absolute magnitude of the cross section, but not the dependence on the collision energy.
Second, a broad resonance which appears at $E_c\sim 2$ cm$^{-1}$ for zero electric field is completely suppressed 
at $E=200$ kV/cm. As we demonstrated earlier for spin-changing collisions of CaD molecules \cite{DCS},
electric fields induce the off-diagonal $\ell \to \ell \pm 1$ transitions, which alters the relative contribution of different
partial waves to the total cross section and suppresses shape resonances \cite{OurWork,DCS}.

An additional feature of the LiF + H reaction is the presence of rovibrational relaxation channels
LiF$(v=1,j=0)$~+~H $\to$ LiF$(v'=0,j')$~+~H competing with the chemical reaction. The lower panel of Fig. 3 shows that the
vibrational relaxation is more probable than the chemical reaction at low collision energies.
This result is consistent with the calculations of Weck and Balakrishnan \cite{BalaLiF}.
We note that because of the shape resonance at $\sim$2 K, the cross section for vibrational relaxation displays the same energy
dependence as the reactive scattering cross section for collision energies above 1 K.
The coupling between different partial waves induced by the electric field modifies the shape resonance and suppresses
both the inelastic and reactive cross sections in the multiple partial-wave regime. Thus, electric fields can be used not only
to change the absolute magnitude of the cross sections,  but also to modify their dependence on the collision energy. 
We note that $s$-wave scattering begins to dominate the cross sections 
for vibrational relaxation at higher collision energies, especially at 
zero electric field where the reaction probabilities are very small. The 
presence of electric fields enhances the reaction  probabilities in the 
limit of $s$-wave scattering, which shifts the upturn of the total reaction 
cross sections to higher collision energies. 
The lower panel of Fig.~3 shows that electric fields stimulate vibrational relaxation in the $s$-wave regime, although the effect is
not as significant as for the chemical reaction.

To elucidate the effects of electric fields on vibrational relaxation, we calculated the cross sections
for the $|v=1,\tau=0\rangle \to |v'=0\rangle$ transition in a simpler non-reactive collision system CaD~+~He.
The calculations were performed as described in Ref. \cite{OurWork}. Figure 4 shows that electric fields stimulate
vibrationally inelastic scattering, in agreement with the results of the calculations for the LiF~+~H reaction (Fig. 3).
The efficiency of vibrational relaxation is determined by the coupling of the ground
rotational state of the $v=1$ level with the rotational states in the $v=0$ manifiold induced by the anisotropy
of the interaction potential. In the absence of an electric field, the dominant transition $|v=1,j=0\rangle \to |v'=0,j'=1\rangle$
is induced by the leading anisotropic term $V_1$ of the Legendre expansion of the interaction potential
$$V(R,r,\gamma)=\sum_\lambda V_\lambda (R,r) P_\lambda(\cos\gamma).$$
Electric fields couple the initial state $|v=1,j=0\rangle$ with the rotationally excited states. As a result, more rotational levels
within the $v=1$ manifold are coupled by the $V_1$ term and the coupling between the ground and the first excited vibrational
states increases. To verify this mechanism, we calculated the cross sections for vibrational relaxation without the $|v=1,j=1\rangle$
level in the basis set.
Figure 4 shows that the cross sections calculated with the modified basis do not change with increasing electric field. This indicates
that the coupling between the ground and the first excited rotational states plays a key role in stimulating vibrationally inelastic
collisions with electric fields. The reactive scattering cross sections display a similar behavior (see Fig. 3), which suggests
that the mechanisms for the electric field enhancement of chemical reactions and vibrationally inelastic collisions are similar.

Figure 5 illustrates that the rate constants for the LiF + H $\to$ HF + Li reaction increase by several orders
of magnitude with increasing field. The increase is not monotonic: the cross section for $E=150$ kV/cm
is one order of magnitude lager than for $E=200$ kV/cm. This indicates the presence of an electric-field
induced resonance similar to those observed by Avdeenkov and Bohn \cite{AvdeenkovBohn} and in our
previous work \cite{PRA} for non-reactive scattering. This new class of reactive scattering resonances is very interesting as it might allow
for controlling chemical reactions with external fields.

Figure 6 shows the total cross section for the Li + HF $\to$ LiF + H reaction as a function of the collision energy and the
electric field strength. The cross sections in the $s$-wave regime ($E_c\sim 1$ mK) are almost unaffected by electric fields.
The effects of electric fields are more pronounced near a scattering resonance at 0.5 K, where the cross sections are
suppressed by a factor of 10. As discussed above, this suppression occurs as a result of the electric field-induced mixing
of different  partial waves \cite{DCS}.
The reaction probability at low temperatures is determined by tunneling under the reaction barrier
\cite{barrier} and the relative change of the reaction exothermicity induced by electric fields
is small (see Fig. 2). The effect of electric fields on the structure of HF is relatively weak
as the dipole moment of HF is quite small and the rotational constant of 
the molecule is large.  We conclude that the total reaction probability is insensitive to the electric-field-induced 
interactions in the exit reaction channel. 
This is in contrast with the effects of electric fields in the entrance reaction channel discussed above.

\section{Summary and conclusions}

In summary, we have developed a quantum mechanical theory of reactive scattering in the presence of an external electric field.
The approach is based on the FD hyperspherical coordinates and the adiabatic partitioning of the reactive scattering Hamiltonian.
The total wave function is expanded in the eigenfunctions of the adiabatic Hamiltonian, which includes the molecule-field
interactions. The expansion coefficients as functions
of the propagation variable $\rho$ are determined from the solution of the coupled-channel equations.
The adiabatic eigenfunctions are constructed on a grid of $\rho$ sectors by solving the adiabatic eigenvalue problem (\ref{AEP}).
The fully uncoupled SF basis of spherical harmonics and hyperspherical ro-vibrational basis functions in each chemical arrangement
is used to expand the adiabatic eigenfunctions (Sec. II C).

The matrix elements of the field-independent part of the adiabatic Hamiltonian (\ref{Had}) are first computed in the total
angular momentum representation using the BF basis functions (\ref{BFbasis}). The integrals are then transformed to the SF
representation using Eq. (\ref{SFBF}). The matrix elements of the interaction with electric fields are computed directly in the SF frame
[Sec. II D, Eq. (\ref{HefJacobi})] and added to the field-independent part (\ref{potential2}) to yield the matrix of the adiabatic Hamiltonian
in the SF primitive basis [Eq. (\ref{MatrixElementsHrho}), right-hand side].
The dipole moment of the atom-molecule system can be represented as a sum of the dipole moments of
the individual diatomic molecules and a three-body term, which vanishes as the atom-molecule separation increases.
The three-body contribution is negligible outside the strong potential coupling region. Inside this region, the interaction
of the three-body term with electric fields is small compared to the atom-molecule interaction.  Neglecting this term
is therefore equivalent to ignoring the coupling between different arrangements
due to the electric field (Sec. II D), and is a good approximation \cite{Vigue}.

The matrix of the adiabatic Hamiltonian is transformed to the orthogonalized basis using Eq. (\ref{MatrixElementsHrho})
and subsequently diagonalized
to yield the adiabatic eigenvalues and eigenfunctions at each $\rho$. During the propagation from small values of $\rho$ to
the asymptotic region,
the wave function (or its logarithmic derivative) is transformed from one sector to another using the sector-to-sector transformation
matrix (\ref{S}). At the end of the propagation, the matrix of solutions is transformed to the asymptotic
basis which diagonalizes the interaction with electric fields (\ref{AsymptoticFunctions}).
The asymptotic matching procedure described in the Appendix yields
the reactance matrix $K$ and the scattering matrix $S$ as well as the integral cross sections for reactive scattering as functions of
the collision energy and the electric field strength (Sec.~II~F).

The theory presented in this work is completely general and can be applied to any abstraction reaction involving
polar molecules. Because the field-independent Hamiltonian (\ref{Had}) remains the
same for any atom-diatom chemical reaction \cite{note1}, the equations derived in Sec. II provide a general framework for
including the effects of external electromagnetic fields in reactive scattering calculations. For example, it is straightforward
to modify the presented formalism to describe chemical reactions in the presence of an off-resonant laser light \cite{Seideman} by
changing the matrix elements of the molecule-field interaction (Sec. II E). Our time-independent approach can be generalized to
study chemical reactions in the presence of microwave laser radiation \cite{MW} or radio-frequency fields \cite{Verhaar} using the
dressed-state formalism \cite{CohenTannoudji,Julienne}. Some of these generalizations are currently under development in our group.
We note that the theory presented here may not be applicable to
insertion reactions involving the formation of long-lived intermediate 
complexes because the FD hyperspherical expansions are known to converge 
very slowly for this type of reactions.  One possible extension of the 
present work would be to develop a formalism based on symmetric  
hyperspherical coordinates of Smith and Whitten \cite{Smith}. This would allow for 
efficient numerical calculations of probabilities for insertion chemical 
reactions in the presence of external fields.

Based on the theory developed Sec. II, we performed preliminary calculations of cross sections and rate constants for
the LiF($v=1,j=0$) + H and
Li + HF($v=0,j=0$) chemical reactions in the presence of an external electric field.  Our calculations show that
cross sections (Fig. 3) and rate constants (Fig. 5) for chemical reactions at low temperatures may be sensitive
to dc electric fields of less than
200 kV/cm. Our results indicate that the probability for the LiF + H reaction in the $s$-wave regime is enhanced dramatically
by moderate electric fields in the range 100 - 200 kV/cm (Fig. 3). Low-temperature rate constants
of abstraction reactions such as LiF + H are typically very small, and the results shown in Fig. 5 suggest that electric fields can be used to
stimulate chemical reactions in cold trapped molecular ensembles. The increase of the rate with electric field is not monotonic,
which suggests that chemical reactions at low temperatures might be affected by Feshbach resonances sensitive to external fields.
We will explore the effects of external fields on scattering resonances in chemical reactions at low temperatures in future work.

Our analysis shows that the enhancement of reaction rates at low temperatures is due to electric field-induced couplings
between rotational states of the reactants. This suggests that chemical reactions at low temperatures can be selectively
tuned by microwave laser fields.
Microwave radiation couples different rotational states of the reactants, thereby inducing coupling between different vibrational states
and reaction channels. 
The tunability and high power of microwave lasers may allow for high selectivity of control. 
A major thrust of recent experimental work has been to produce cold and dense ensembles of a variety of stable molecular
radicals such as NH \cite{NNH,Wes}, CaH \cite{Weinstein}, and OH \cite{StarkDeceleration,MeijerXeOH}.
The mechanisms for the electric field control of chemical reactions  described in this work can be
readily verified in the experimental work with slow molecular beams \cite{StarkDeceleration,MeijerXeOH,Rempe}
or with molecules confined in a magnetic or electrostatic trap
\cite{Rempe,MeijerXeOH,NNH}. The reaction products in external field traps could be separated and detected independently
using the $E-H$ gradient balance method \cite{EH}.

\acknowledgements

We thank David Manolopoulos for stimulating discussions. This work was supported by the Killam Trusts and NSERC of Canada.
The allocation of computer time on Western Canada Research Grid (Westrid) is gratefully acknowledged.

\appendix
\section{Asymptotic Matching}

Because the asymptotic form of the Jacobi radial functions is well known (\ref{F}), it is convenient 
to re-express the FD wave function at $\rho\to \infty$  in the Jacobi coordinates (\ref{SpaceFixedJacobi}).
The asymptotic FD wave function at the end of the propagation after the transformation
to the asymptotic basis (see Sections IID and IIF) has the form
\begin{equation}\label{ExpansionFD}
\Psi = \sum_{\alpha, v,\tau,M_\tau}\sum_{\l, M_\l} \rho^{-5/2} F_{\alpha v\tau M_\tau \l M_\l}(\rho)
\psi_{\alpha v\tau M_\tau}(\hr,\theta_\alpha;\rho) Y_{\l M_\l}(\hR),
\end{equation}
where the field-dressed FD pendular states $\psi_{\alpha v\tau M_\tau}(\hr,\theta_\alpha;\rho) $ are given by Eq. (\ref{AF}).
Note that the field-dependent expansion coefficients $C_{jM_j, \tau M_\tau}$ are the same as in Eq. (\ref{A1}) because
the matrix elements of the asymptotic Hamiltonian are identical in the FD and Jacobi coordinates (see Sec. II D).

Since the basis functions of different arrangements are orthonormal in the limit of large $\rho$, we can invert
Eq. (\ref{ExpansionFD}) to obtain
\begin{equation}
F_{\alpha v\tau M_\tau \l M_\l}(\rho) = \int d\hr \int d\hR \int_0^{\pi/2} d\theta_\alpha \textstyle{\frac{1}{4}}\sin^2 2\theta_\alpha
(\rho^{5/2} \Psi)
\psi^*_{\alpha v\tau M_\tau}(\hr,\theta_\alpha;\rho) Y^*_{\l M_\l}(\hR),
\end{equation}
where $\frac{1}{4}\sin^2 2\theta_\alpha$ is the angular part of the Jacobian in the FD coordinates \cite{ParkerPack,Review}.
Substituting $\Psi$ from Eq. (\ref{SpaceFixedJacobi}) yields
\begin{multline}\label{Projection1}
F_{\alpha v\tau M_\tau \l M_\l}(\rho) = \int d\hr \int d\hR \int_0^{\pi/2} d\theta_\alpha \textstyle{\frac{1}{4}}\sin^2 2\theta_\alpha
\psi^*_{\alpha v\tau M_\tau}(\hr,\theta_\alpha;\rho) Y^*_{\l M_\l}(\hR) \\ \times
\sum_{\alpha' v'\tau'M_\tau'}\sum_{\l', M_\l'} \frac{\rho^{5/2}}{R_{\alpha'}r_{\alpha'}}F_{\alpha v\tau M_\tau \l M_\l}(R_{\alpha'})
\psi_{\alpha' v'\tau' M_\tau'}({\bf r}_{\alpha'})Y_{\l' M_\l'}(\hat{R}_{\alpha'}).
\end{multline}
As mentioned above, in the limit of large $\rho$ the basis functions have zero overlap unless $\alpha=\alpha'$.
Using the definition (\ref{A1}) and the orthogonality property of the coefficients $C_{jM_j,\tau M_\tau}$,
the integration over $\hat{r}_\alpha$ can be carried out analytically to yield
\begin{equation}
\int d\hr
\psi^*_{\alpha v\tau M_\tau}(\hr,\theta_\alpha;\rho) \psi_{\alpha v'\tau' M_\tau'}({\bf r}_{\alpha})
=\delta_{\tau\tau'}\delta_{M_\tau M_\tau'}\chi_{\alpha v\tau}(\theta_\alpha;\rho)\xi_{\alpha v'\tau}(r_\alpha).
\end{equation}
The integration over $\hR$ is straightforward because the expansions given by Eqs. (\ref{SpaceFixedJacobi}) and (\ref{ExpansionFD})
contain the same spherical harmonics. The matrix element in Eq. (\ref{Projection1}) 
reduces after some algebra to the sum of one-dimensional integrals \cite{ParkerPack,Review}
\begin{multline}\label{Projection2}
F_{\alpha v\tau M_\tau \l M_\l}(\rho) = 
\sum_{\alpha' v'\tau'M_\tau'}\sum_{\l', M_\l'} \delta_{\alpha\alpha'}\delta_{\tau \tau'}\delta_{M_\tau M_\tau'}
\delta_{\l \l'}\delta_{M_\l M_\l'} \\ \times
 \int_0^{\pi/2} d\theta_\alpha \phi_{\alpha v\tau}(\theta_\alpha;\rho)  F_{\alpha' v'\tau' M_\tau' \l' M_\l'}(R_{\alpha'})
 \xi_{\alpha' v'\tau'}(r_{\alpha'}),
\end{multline}
where the renormalized FD rovibrational function $\phi_{\alpha v\tau}(\theta_\alpha;\rho) $ is given by Eq. (\ref{Renormalized}).

Instead of Eq. (\ref{F}), it is convenient to use the real boundary conditions for the Jacobi radial functions 
\begin{equation}\label{JN}
\mathsf{F}(R_\alpha) = \mathsf{J}(R_\alpha) - \mathsf{N}(R_\alpha)\mathsf{K},
\end{equation}
and their derivatives
\begin{equation}\label{JND}
\mathsf{F}'(R_\alpha) = \mathsf{J}'(R_\alpha) - \mathsf{N}'(R_\alpha)\mathsf{K},
\end{equation}
where the primes indicate differentiation with respect to $R_\alpha$ in arrangement $\alpha$. In Eq. (\ref{JN}),
$\mathsf{K}$ is the reactance $K$-matrix, and the diagonal matrices of incoming and outgoing waves
$\mathsf{J}(R)$ and $\mathsf{N}(R)$
are composed of the modified spherical Bessel functions \cite{Review,ParkerPack,Johnson}. Following
Parker and Pack \cite{ParkerPack}, we substitute Eqs. (\ref{JN}) and (\ref{JND}) into Eq. (\ref{Projection2}) to obtain
\begin{equation}\label{FAB}
\mathsf{F}(\rho) = \mathsf{A}(\rho) - \mathsf{B}(\rho)\mathsf{K},
\end{equation}
where
\begin{multline}\label{AB}
[\mathsf{A}(\rho)]_{\alpha v \tau M_\tau \l M_\l, \alpha'v'\tau' M_\tau' \l' M_\l'} = 
\delta_{\alpha\alpha'}\delta_{\tau \tau'}\delta_{M_\tau M_\tau'}
\delta_{\l \l'}\delta_{M_\l M_\l'} \rho^{1/2} \\ \times
\int_0^{\pi/2} d\theta_\alpha \phi_{\alpha v\tau}(\theta_\alpha;\rho)
[\mathsf{J}(R_{\alpha'})]_{\alpha' v'\tau' M_\tau' \ell' M_\ell'} \xi_{\alpha' v'\tau'}(r_{\alpha'}),
\end{multline}
The matrix $\mathsf{B}(\rho)$ has the same form with the Bessel functions $\mathsf{J}(R_\alpha)$
substituted by $\mathsf{N}(R_\alpha)$
Taking the first derivative of Eq. (\ref{Projection2}) with respect to $\rho$, we obtain
\begin{equation}\label{FAB}
\mathsf{F}'(\rho) = \frac{1}{2\rho}\mathsf{F}(\rho) \mathsf{A}(\rho) +\mathsf{G}(\rho) - \mathsf{H}(\rho)\mathsf{K},
\end{equation}
where the matrix $\mathsf{G}$ is given by
\begin{multline}\label{GH}
[\mathsf{G}(\rho)]_{\alpha v \tau M_\tau \l M_\l, \alpha'v'\tau' M_\tau' \l' M_\l'} = 
\delta_{\alpha\alpha'}\delta_{\tau \tau'}\delta_{M_\tau M_\tau'}
\delta_{\l \l'}\delta_{M_\l M_\l'} \rho^{1/2}
\int_0^{\pi/2} d\theta_\alpha \phi_{\alpha v\tau}(\theta_\alpha;\rho) \\ \times
\left[ \frac{d[\mathsf{J}(R_{\alpha'})]_{\alpha' v'\tau' M_\tau' \ell' M_\ell'}}{dR_{\alpha'}}
\xi_{\alpha'v'\tau'}(r_{\alpha'}) \cos\theta_{\alpha'}+ 
[\mathsf{J}(R_{\alpha'})]_{\alpha' v'\tau' M_\tau' \ell' M_\ell'} \frac{d\xi_{\alpha'v'\tau'}(r_{\alpha'})}{dr_{\alpha'}}\sin\theta_{\alpha'}   \right],
\end{multline}
and the matrix $\mathsf{H}(\rho)$ has the identical form, with the Bessel functions $\mathsf{J}(R_\alpha)$
substituted by $\mathsf{N}(R_\alpha)$. From Eqs. (\ref{AB}) and (\ref{GH}), we can obtain the K-matrix
in the Jacobi coordinates directly from the log-derivative matrix \cite{Johnson,Manolopoulos,ParkerPack,Review}
in the asymptotic FD basis
\begin{equation}\label{K}
\mathsf{K} = \left[ \left( \mathsf{Y} - \frac{1}{2\rho}\mathsf{I} \right)\mathsf{B} - \mathsf{H}\right]^{-1}
\left[ \left( \mathsf{Y} - \frac{1}{2\rho}\mathsf{I} \right)\mathsf{A} - \mathsf{G}\right],
\end{equation}
where $\mathsf{I}$ is the identity matrix. The $S$-matrix can be obtained from the $K$-matrix
using the standard Cayley transformation \cite{ParkerPack,Review}.
As in the conventional reactive scattering formalism \cite{ParkerPack,ABC,Review}, the projection matrices
(\ref{AB}) and (\ref{GH}) are diagonal in all quantum numbers except $v$.

\newpage

\begin{figure}
	\centering
	\includegraphics[width=0.7\textwidth, trim = 0 00 0 60]{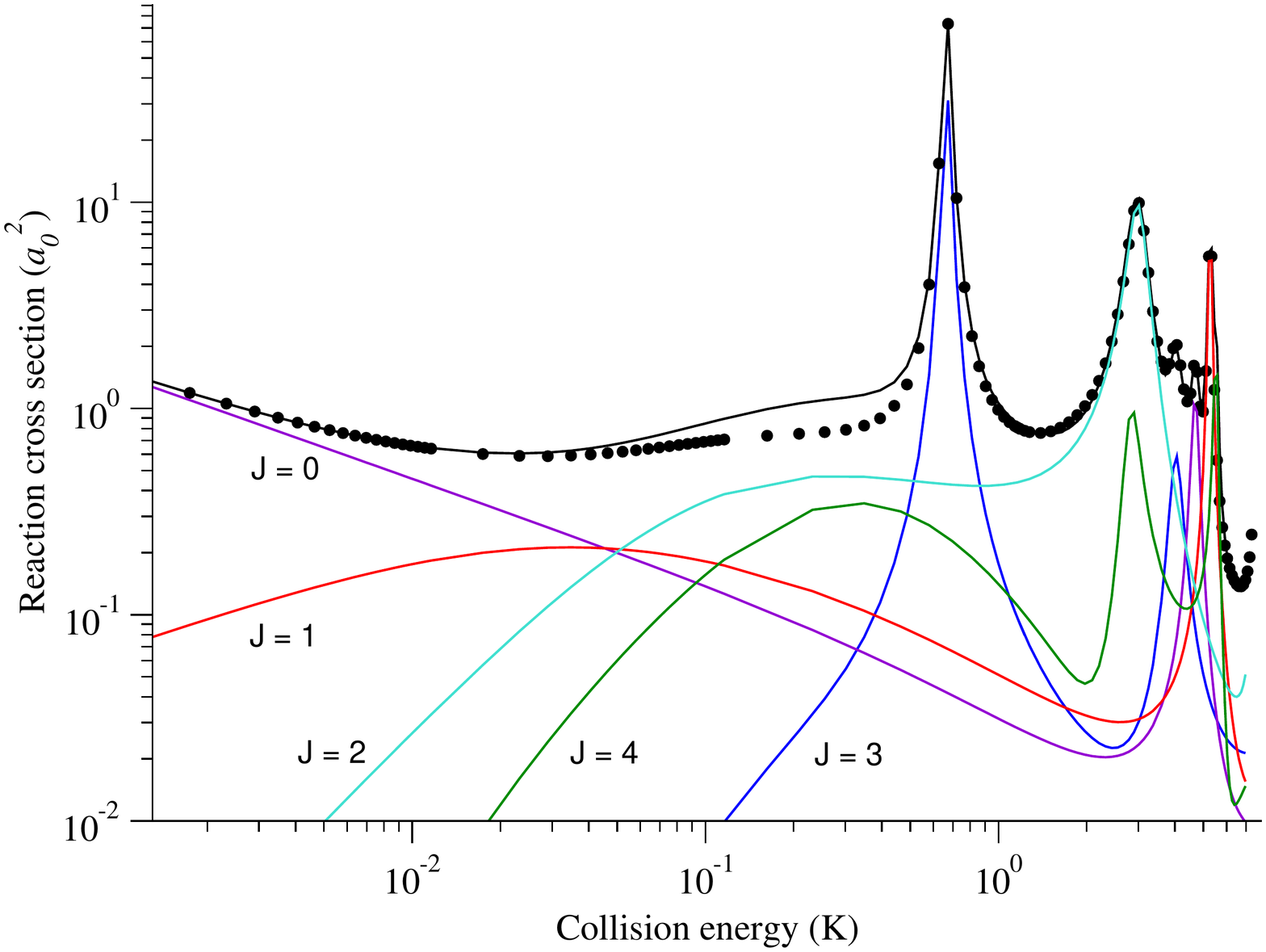}
	\renewcommand{\figurename}{Figure}
	\caption{Total cross sections for the Li + HF $\to$ LiF + H reaction calculated using the SF uncoupled representation (circles) and the
	conventional BF method (full line). The SF cross sections are computed using Eq. (\ref{ICS}) and summed over $\ell=0-5$.
	The BF cross sections are computed as $\sigma = \sum_J (2J+1) \sigma^J$ and summed over $J=0-4$.
	The partial BF cross sections [$(2J+1)\sigma^J$] are also presented.}
\end{figure}

\begin{figure}
	\centering
	\includegraphics[width=0.65\textwidth, trim = 0 20 0 60]{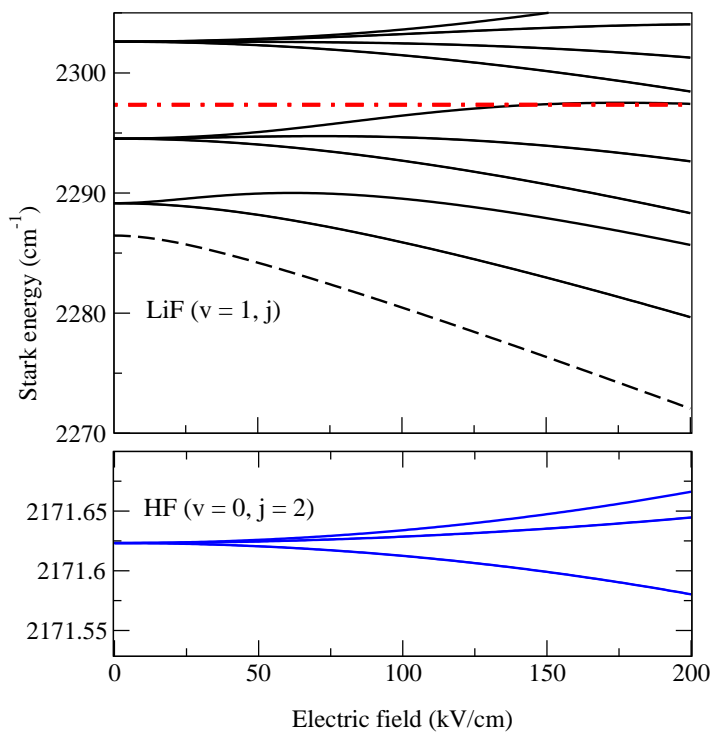}
	\renewcommand{\figurename}{Figure}
	\caption{Energy levels of the reactants and products of the LiF($v=1,j=0$) + H $\to$ HF + Li reaction as functions of the electric field.
	The initial level $|v=1,j=0\rangle$ of LiF is shown by the dashed line. The HF states correlating to
	the $|v=0,j=3\rangle$ level in the zero-field limit are shown by the dash-dotted line in the upper panel.
	The zero of energy corresponds to the bottom of the LiHF potential well \cite{Werner} at zero electric field.}
\end{figure}

\begin{figure}
	\centering
	\includegraphics[width=0.65\textwidth, trim = 0 20 0 60]{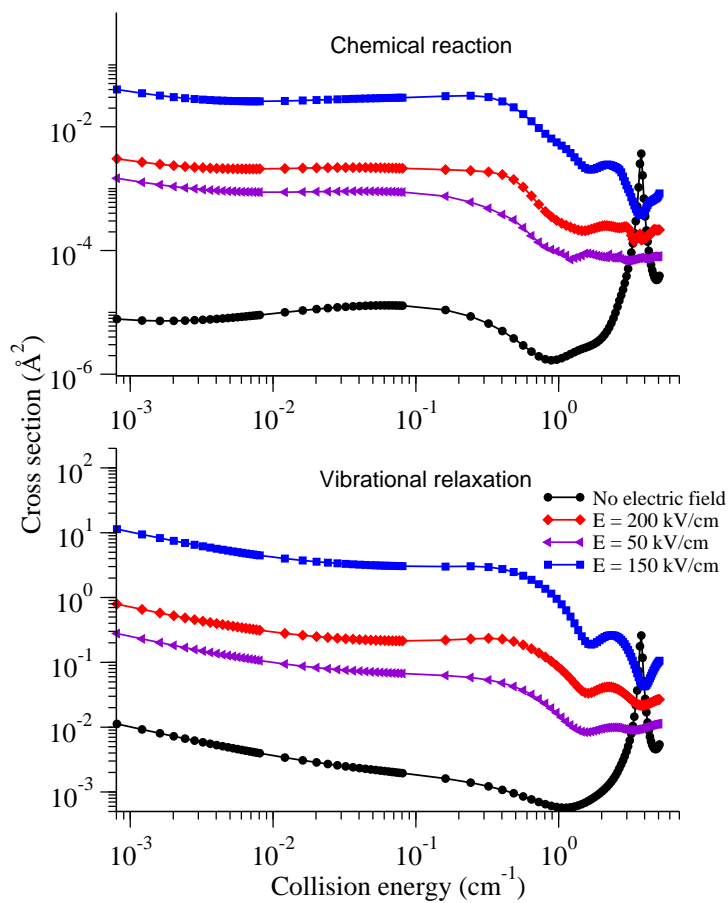}
	\renewcommand{\figurename}{Figure}
	\caption{Upper panel: Total cross sections for the LiF($v=1,j=0$) + H chemical reaction as functions of the collision energy at zero electric
	field (circles), $E$ = 100 kV/cm (triangles), E = 150 kV/cm (squares), and $E=200$ kV/cm (diamonds). Lower panel:
	Cross sections for vibrational relaxation LiF($v=1,j=0$) + H $\to$ LiF($v'=0$) + H summed over all final field-dressed
	rotational states as functions of the collision energy at different electric field strengths.}
\end{figure}

\begin{figure}
	\centering
	\includegraphics[width=0.65\textwidth, trim = 0 60 0 60]{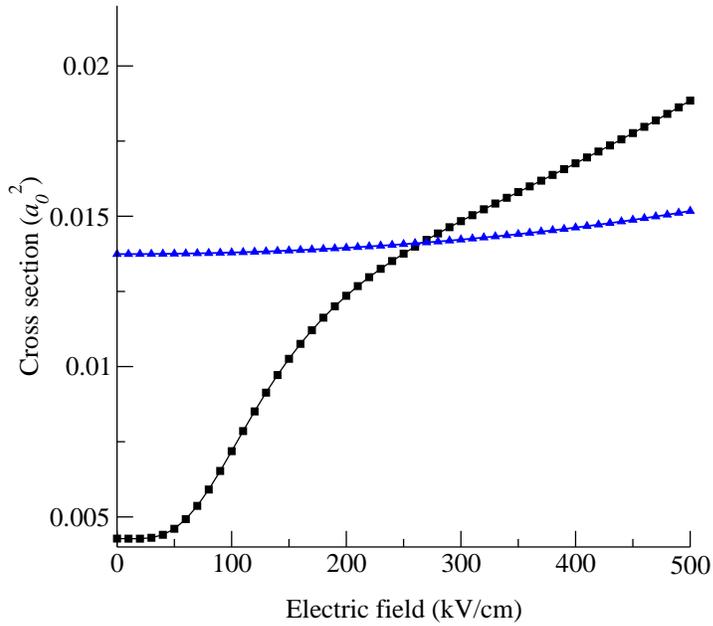}
	\renewcommand{\figurename}{Figure}
	\caption{Cross sections for vibrational relaxation $|v=1,j=0\rangle \to |v'=0\rangle$ in CaD + He collisions as functions of the
	applied electric field: fully converged calculations (squares), and the results obtained without the $j=1$ level (triangles).	 The cross sections are calculated for $M=0$ and summed over all final field-dressed rotational states. The collision energy is $10^{-3}$ K.}
\end{figure}

\begin{figure}
	\centering
	\includegraphics[width=0.65\textwidth, trim = 0 20 0 60]{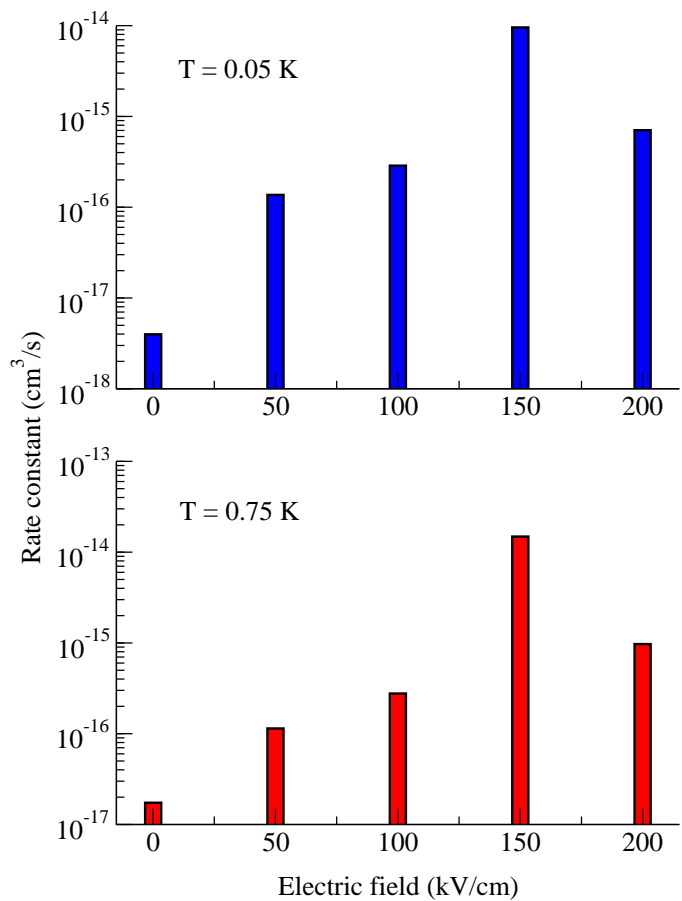}
	\renewcommand{\figurename}{Figure}
	\caption{Rate constants for the LiF($v=0,j=1$) + H $\to$ HF + Li reaction at two different temperatures
	as functions of the applied electric field. Upper panel: $T=0.05$ K; Lower panel: $T=0.75$ K.}
\end{figure}

\begin{figure}
	\centering
	\includegraphics[width=0.65\textwidth, trim = 0 60 0 60]{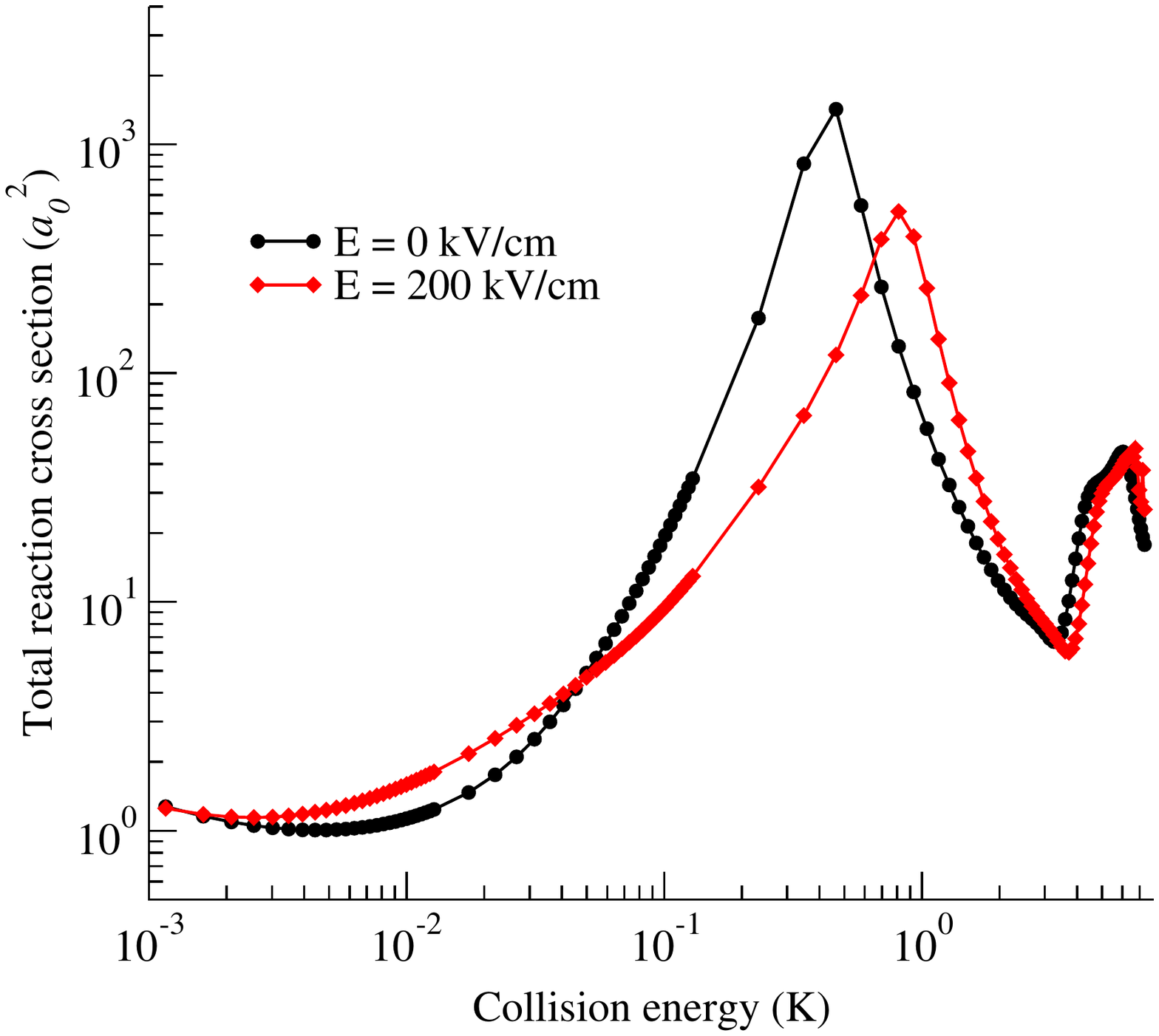}
	\renewcommand{\figurename}{Figure}
	\caption{Total integral cross sections for the Li + HF reaction as functions of the collision energy at
	zero electric field (circles) and $E=200$ kV/cm (diamonds).}
\end{figure}


\newpage
\begin{thebibliography}{99}

\bibitem{Zare}
R. N. Zare, Science {\bf 279}, 1875 (1998).

\bibitem{Roman}
R. V. Krems, Int. Rev. Phys. Chem. {\bf 24}, 99 (2005); J. Doyle, B. Friedrich, R. V. Krems,
and F. Masnou-Seeuws, Eur. Phys. J. D {\bf 31}, 149 (2004).

\bibitem{LoeschReview}
H. J. Loesch and A. Remschield, \jcp{93}{4779}{1990};
H.-J. Loesch, Annu. Rev. Phys. Chem. {\bf 46}, 555 (1995).

\bibitem{Stereodynamics}
V. Aquilanti, M. Bartolomei, F. Pirani, D. Cappelletti, F. Vecchiocattivi, Y. Shimizu, and T. Kasai,
Phys. Chem. Chem. Phys. {\bf 7}, 291 (2005).


\bibitem{EH}
T. J. McCarthy, M. T. Timko, and D. R. Herschbach, \jcp{125}{133501}{2006}.


\bibitem{Moshe}
M. Shapiro and P. Brumer, {\it Principles of Quantum Control of Molecular Processes} (Wiley, NJ, 2003);
S. A. Rice and M. Zhao, {\it Optical Control of Molecular Dynamics} (Wiley, NY, 2000).

\bibitem{Bretislav}
B. Friedrich, M.-G. Rubahn, and N. Sathyamurthy, \prl{69}{2487}{1992}.



\bibitem{Steric1}
R. J. Beuhler, Jr., R. B. Bernstein, and K. H. Kramer, J. Am. Chem. Soc. {\bf 88}, 5331 (1966).


\bibitem{Brooks}
P. R. Brooks, Science {\bf 193}, 11 (1976).

\bibitem{Stolte}
H. Thuis, S. Stolte, and J. Reuss, \cpl{43}{351}{1979};
D. van den Ende and S. Stolte, \cpl{76}{13}{1980};
D. van den Ende and S. Stolte, Chem. Phys. {\bf 89}, 121 (1984).


\bibitem{Loesch2}
H.-J. Loesch and F. Stienkemeier, \jcp{98}{9570}{1993}.

\bibitem{LoeschMoller}
H. J. Loesch and J. M{\"o}ller, \jcp{97}{9016}{1992}.


\bibitem{Bretislav1}
B. Friedrich and D. Herschbach, Nature (London) {\bf 353}, 412 (1991);
B. Friedrich, A. Slenczka, and D. Herschbach, Can. J. Phys. {\bf 72}, 897 (1994).

\bibitem{FH1995}
B. Friedrich and D. Herschbach, \prl{74}{4623}{1995}.

\bibitem{Seideman}
H. Stapelfeldt and T. Seideman, Rev. Mod. Phys. {\bf 75}, 543 (2003).

\bibitem{Stapelfeldt}
J. J. Larsen, H. Sakai, C. P. Safvan, I. Wendt-Larsen, and H. Stapelfeldt,
\jcp{111}{7774}{1999}.

\bibitem{Corkum}
J. Levesque, Y. Mairesse, N. Dudovich, H. P\'epin, J.-C. Kieffer, P. B. Corkum,
and D. M. Villeneuve, \prl{99}{243001}{2007}.

\bibitem{Stapelfeldt2}
S. S. Viftrup, V. Kumarappan, S. Trippel, and H. Stapelfeldt, \prl{99}{143602}{2007}.

\bibitem{Stapelfeldt3}
J. J. Larsen, I. Wendt-Larsen, and H. Stapelfeldt, \prl{83}{1123}{1999}.

\bibitem{CollisionalA}
D. P. Pullman, B. Friedrich, and D. R. Herschbach, \jcp{93}{3224}{1990};
M. J. Wieda and D. J. Nesbitt, \jcp{100}{6372}{1994}.

\bibitem{RomanPCCP}
R. V. Krems, Phys. Chem. Chem. Phys., in press (2008).


\bibitem{BalaReview}
P. F. Weck and N. Balakrishnan, Int. Rev. Phys. Chem. {\bf 25}, 283 (2006).

\bibitem{BalaLiF}
P. Weck and N. Balakrishnan, \jcp{122}{234310}{2005}.


\bibitem{UltracoldChemistry}
N. Balakrishnan and A. Dalgarno, Chem. Phys. Lett. {\bf 341}, 652 (2001);
E. Bodo, F. A. Gianturco, and A. Dalgarno, \jcp{116}{9222}{2002};
E. Bodo, F. A. Gianturco, N. Balakrishnan, and A. Dalgarno, \jpb{37}{3641}{2004}.


\bibitem{SuperChemistry}
D. J. Heinzen, R. Wynar, P. D. Drummond, and K. V. Kheruntsyan, \prl{84}{5029}{2000}.


\bibitem{NatureInsight}
Nature Insight: Ultracold Matter, Nature (London) {\bf 416}, 205 (2002).

\bibitem{StarkDeceleration}
S. Y. T. van de Meerakker, N. Vanhaecke, and G. Meijer, Annu. Rev. Phys. Chem. {\bf 57}, 159 (2006).


\bibitem{MeijerXeOH}
J. J. Gilijamse, S. Hoekstra, S. Y. T. van de Meerakker, G. C. Groenenboom, and G. Meijer,
Science {\bf 313}, 1617 (2006).


\bibitem{Rempe}
M. Motsch, M. Schenk, L. D. van Buuren, M. Zeppenfeld, P. W. H. Pinkse, and G. Rempe,
\pra{76}{061402(R)}{2007}; D. Patterson and J. M. Doyle, \jcp{126}{154307}{2007}.


\bibitem{MeijerMS}
C. E. Heiner, D. Carty, G. Meijer, and H. L. Bethlem, Nat. Phys. {\bf 3}, 115 (2007); R. V. Krems, Nat. Phys. {\bf 3}, 77 (2007).


\bibitem{NNH}
M. T. Hummon, W. C. Campbell, H.-I Lu, E. Tsikata, Y. Wang, and J. M. Doyle, preprint atom-ph/0802.1662 (2008).

\bibitem{Alkali1}
P. Staanum, S. D. Kraft, J. Lange, R. Wester, and M. Weidem{\"u}ller,
\prl{96}{023201}{2006}.

\bibitem{Alkali2}
N. Zanzam, T. Vogt, M. Mudrich, D. Comparat, and P. Pillet,
\prl{96}{023202}{2006}.


\bibitem{HudsonFormaldehyde}
E. R. Hudson, C. Ticknor, B. C. Sawyer, C. A. Taatjes, H. J. Lewandowski, J. R. Bochinski, J. L. Bohn, and J. Ye,
 \pra{73}{063404}{2006}.

\bibitem{HeraeusSeminar}
M. Mudrich, in 379. WE-Heraeus-Seminar on Cold Molecules (2006).




\bibitem{Steric2}
M. Karplus and M. Godfrey, J. Am. Chem. Soc. {\bf 88}, 5332 (1966).

\bibitem{BretislavQCT}
F. J. Aoiz, B. Friedrich, V. J. Herrero, V. S{\'a}ez-R{\'a}banos, and J. E. Verdasco, \cpl{289}{132}{1998}.

\bibitem{Aldegunde1}
J. Aldegunde, M. P. de Miranda, J. M. Haigh, B. K. Kendrick, V. S{\'a}ez-R{\'a}banos, F. J. Aoiz,
J. Phys. Chem. A {\bf 109}, 6200 (2005).

\bibitem{Aldegunde2}
J. Aldegunde, J. M. Alvari$\tilde{\text{n}}$o, M. P. de Miranda, V. S\'aez Rabanos, F. J. Aoiz,
\jcp{125}{133104}{2006}.

\bibitem{Enzo}
J. M. Alvari$\tilde{\text{n}}$o, V. Aquilanti, S. Cavalli, S. Crocchianti, A. Lagan{\`a}, and T. Mart{\'i}nez,
\jcp{107}{3339}{1997}; J. Aldegunde, J. M. Alvari$\tilde{\text{n}}$o, D. De Fazio, S. Cavalli, G. Grossi,
and V. Aquilanti, Chem. Phys. {\bf 301}, 251 (2004).

\bibitem{OM}
A. E. Orel and W. H. Miller, \jcp{70}{4393}{1979}; \jcp{73}{241}{1980}.

\bibitem{Light}
J. C. Light and A. Altenberger-Siczek, \jcp{70}{4108}{1979}.

\bibitem{Ivanov}
J. C. Peploski and L. Eno, \jcp{83}{2947}{1985};
D. R. Matusek, M. Yu. Ivanov, and J. S. Wright, \cpl{258}{255}{1996}.


\bibitem{Moshe1}
T. Seideman and M. Shapiro,
J. Chem. Phys. {\bf 94}, 7910 (1991).

\bibitem{Moshe2}
J. L. Krause and M. Shapiro, \jcp{92}{1126}{1990}.

\bibitem{Moshe3}
X. Li, G. A. Parker, P. Brumer, I. Thanopulos, and M. Shapiro,
\jcp{128}{124314}{2008}.

\bibitem{TannorRice}
D. J. Tannor and S. A. Rice, \jcp{83}{5013}{1985};
D. J. Tannor, R. Kosloff, and S. A. Rice, \jcp{85}{5805}{1986};
R. Kosloff, S. A. Rice, P. Gaspard, S. Tersigni, and D. J. Tannor,
Chem. Phys. {\bf 139}, 201 (1989).





\bibitem{Volpi}
A. Volpi and J. L. Bohn, Phys. Rev. A {\bf 65}, 052712 (2002).


\bibitem{Roman2004}
R. V. Krems and A. Dalgarno, \jcp{120}{2296}{2004}.

\bibitem{OurWork}
T. V. Tscherbul, and R. V. Krems, \prl{97}{083201}{2006};
T. V. Tscherbul, and R. V. Krems, \jcp{125}{194311}{2006};
E. Abrahamsson, T. V. Tscherbul, and R.~V. Krems, \jcp{127}{044302}{2007}.

\bibitem{DCS}
T. V. Tscherbul, J. Chem. Phys., in press (2008).


\bibitem{ParkerPack}
G. A. Parker and R. T Pack, \jcp{87}{3888}{1987}.

\bibitem{Review}
T. V. Tscherbul and R. V. Krems, manuscript in preparation (2008).

\bibitem{LandauLifshitz}
L. D. Landau and E. M. Lifshits, {\it Quantum Mechanics: Non-Relativistic Theory} (Butterworth-Heineman, 1981).

\bibitem{Tolstikhin}
O. I. Tolstikhin, S. Watanabe, and M. Matsuzawa, J. Phys. B {\bf 29}, L389 (1996).



\bibitem{Truhlar}
M. Topaler, P. Piecuch, and D. Truhlar, \jcp{110}{5634}{1999};
M. Paniagua, A. Aguado, M. Lara, and O. Roncero, \jcp{111}{6712}{1999}.

\bibitem{Vigue}
M. B{\"u}chner, G. Bazalgette, and J. Vigu{\'e}, J. Phys. Chem. A {\bf 101}, 7634 (1997).


\bibitem{Roman2006}
R. V. Krems, \prl{96}{123202}{2006}.


\bibitem{David}
D. E. Manolopoulos, in {\it The Encyclopedia of Computational Chemistry},
Ed. P. von R. Schleyer (Wiley, Chichester, 1998), p. 2699.

\bibitem{HuSchatz}
W. Hu and G. C. Schatz, \jcp{125}{132301}{2006}, and references therein.

\bibitem{NymanYu}
G. Nyman and H.-G. Yu, Rep. Prog. Phys. {\bf 63}, 1001 (2000).


\bibitem{Miller}
J. Z. H. Zhang and W. H. Miller, \jcp{91}{1528}{1989};
W. H. Miller, Annu. Rev. Phys. Chem. {\bf 41}, 245 (1990).


\bibitem{AMW}
M. H. Alexander, D. E. Manolopoulos, and H.-J. Werner, \jcp{113}{11084}{2000}.


\bibitem{AD}
A. M. Arthurs and A. Dalgarno, Proc. R. Soc. London, Ser. A {\bf 256}, 540 (1960).


\bibitem{ABC}
D. Skouteris, J. F. Castillo, and D. E. Manolopoulos, Comp. Phys. Commun. {\bf 133}, 128 (2000).

\bibitem{Lester}
W. A. Lester, in {\it Dynamics of Molecular Collisions}, edited by W. H. Miller (Plenum, New York, 1976).



\bibitem{Szabo}
A. Szabo, N. S. Ostlund, {\it Modern Quantum Chemistry}, (McGraw-Hill, New York, 1989).

\bibitem{Lowdin}
P. O. L{\"o}wdin, \jcp{18}{365}{1950}.


\bibitem{Smith}
F. T. Smith, J. Math. Phys. {\bf 3}, 735 (1962);
R. C. Whitten and F. T. Smith, J. Math. Phys. {\bf 9}, 1103 (1968).


\bibitem{Hutson}
J. M. Hutson, in {\it Advances in Molecular Vibrations and Collision Dynamics}, Eds. J. M. Bowman and M. A. Ratner (JAI press, 1991), p. 1.

\bibitem{ZareBook}
R. N. Zare, {\it Angular Momentum}, (Wiley, New York, 1988).

\bibitem{Varshalovich}
D. A. Varshalovich, A. N. Moskalev, and V. K. Khersonskii, {\it Quantum Theory of Angular Momentum} (World Scientific, New Jersey, 1988).


\bibitem{Johnson}
B. R. Johnson, J. Comp. Phys. {\bf 13}, 455 (1973).

\bibitem{Manolopoulos}
D. E. Manolopoulos, \jcp{85}{6425}{1986}.


\bibitem{ColbertMiller}
D. T. Colbert and W. H. Miller, \jcp{96}{1982}{1992}.

\bibitem{Private}
D. E. Manolopoulos, private communication.

\bibitem{SchatzKuppermann}
G. C. Schatz, \cpl{150}{92}{1988}; S. A. Cuccaro, P. Hipes, and A. Kuppermann, \cpl{154}{155}{1989}.

\bibitem{Werner}
R. Bobbencamp, A. Paladini, A. Russo, H.-J. Loesch, M. Men\'endez, E. Verdasco, F. J. Aoiz,
and H.-J. Werner, \jcp{122}{244304}{2005}.


\bibitem{barrier}
The barrier height for the Li + HF reaction is 1782.5 cm$^{-1}$ \cite{Werner}.

\bibitem{SO2}
S. Jung, E. Tiemann, and C. Lisdat, J. Phys. B {\bf 39}, S1085 (2006);
S. Jung, E. Tiemann, and C. Lisdat, \pra{74}{040701(R)}{2006}.

\bibitem{RomanPRL2004}
R. V. Krems, \prl{93}{013201}{2004}.

\bibitem{DipoleMoments}
L. Wharton, W. Klemperer, L. P. Gold, R. Strauch,  J. J. Gallagher, and V. E. Derr, \jcp{38}{1203}{1963};
A. J. Hebert, F. J. Lovas, C. A. Melendres, C. D. Hollowell, T. L. Story, Jr., and K. Street, Jr., \jcp{48}{2824}{1968};
H.-J. Werner and P. Rosmus, \jcp{73}{2319}{1980}.

\bibitem{AvdeenkovBohn}
A. V. Avdeenkov and J. L. Bohn, \pra{66}{052718}{2002}.

\bibitem{PRA}
T. V. Tscherbul, J. K{l}os, L. Rajchel, and R. V. Krems, \pra{75}{033416}{2007}.

\bibitem{note1}
Chemical reactions of open-shell species may be influenced by non-adiabatic effects
and spin-orbit interactions. Although this suggests new interesting control mechanisms \cite{OurWork}, we do not consider 
these interactions in the present work.

\bibitem{MW}
D. DeMille, D. R. Glenn, and J. Petricka, Eur. Phys. J. D {\bf 31}, 375 (2004).

\bibitem{Verhaar}
A. J. Moerdijk, B. J. Verhaar, and T. M. Nagtegaal, \pra{53}{4343}{1996}.

\bibitem{CohenTannoudji}
C. Cohen-Tannoudji, J. Dupont-Roc, G. Grynberg, {\it Atom-Photon Interactions: Basic Processes and Applications}
(John Wiley and Sons, New York, 1992).

\bibitem{Julienne}
P. S. Julienne, \prl{81}{698}{1988}.

\bibitem{Wes}
W. C. Campbell, E. Tsikata, H. Lu, L. D. van Buuren, and J. M. Doyle, \prl{98}{213001}{2007};
W. C. Campbell, T. V. Tscherbul, H.-I. Lu, E. Tsikata, R. V. Krems, and J. M. Doyle, preprint atom-ph/0804.0265 (2008).

\bibitem{Weinstein}
J. D. Weinstein, R. deCarvalho, T. Guillet, B. Friedrich, and J. M. Doyle, Nature (London) {\bf 395}, 148 (1998).



\end{thebibliography}
\end{document}